\documentclass{aa}
\usepackage[varg]{txfonts}
\usepackage{graphicx}
\usepackage{setspace}
\usepackage{natbib}
\usepackage{longtable}
\usepackage{amssymb}

\bibpunct{(}{)}{;}{a}{}{,}

\begin{document}
\newcommand{\mic} {$\mu$m}
\newcommand{\HI}{\ion{H}{i}}
\newcommand{\brg}{Br$\gamma$}
\newcommand{\NaI}{\ion{Na}{i}}
\newcommand{\macc}{$\dot{M}_{acc}$}
\newcommand{\lacc}{L$_{acc}$}
\newcommand{\lbol}{L$_{bol}$}
\newcommand{\kms}{km\,s$^{-1}$}
\newcommand{\um}{$\mu$m}
\newcommand{\lam}{$\lambda$}
\newcommand{\msyr}{M$_{\odot}$\,yr$^{-1}$}
\newcommand{\Av}{A$_V$}
\newcommand{\Msun}{M$_{\odot}$}
\newcommand{\Lsun}{L$_{\odot}$}
\newcommand{\Rsun}{R$_{\odot}$}
\newcommand{\Lstar}{L$_{\star}$}
\newcommand{\Tstar}{T$_{\star}$} 
\newcommand{\Nco}{$N_{\mathrm{CO}}$} 
\newcommand{\vsini}{$v_{rot}\sin{i}$}
\newcommand{\deltav}{$\Delta \rm v$}
\newcommand {\Tbb} {$T_{\rm BB}$}
\newcommand{\up}{$\upsilon$}
\newcommand{\egs}{erg\,s$^{-1}$\,cm$^{-2}$\,\AA$^{-1}$}
\newcommand{\simless}{\mathbin{\lower 3 pt\hbox {$\rlap{\raise 5pt\hbox{$\char'074$}}\mathchar"7218$}}} 
\newcommand{\simgreat}{\mathbin{\lower 3pt\hbox {$\rlap{\raise 5pt\hbox{$\char'076$}}\mathchar"7218$}}} 
%
\title{The GRAVITY Young Stellar Object survey}
\subtitle{IV. The CO overtone emission in 51\,Oph at sub-au scales}
\author{GRAVITY Collaboration(*): M. Koutoulaki
          \inst{1,2,3,4}
           \and R. Garcia Lopez
           \inst{1,2,3}
           \and A. Natta
           \inst{2}
           \and R. Fedriani
           \inst{2,3,5}
           \and A. Caratti o Garatti
           \inst{1,2,3}
           \and T. P. Ray
           \inst{2}
           \and D. Coffey \inst{2,1}
           \and W. Brandner
           \inst{1}
           \and C. Dougados
           \inst{16}
           \and P.J.V Garcia
           \inst{7,8,11}
           \and L. Klarmann
           \inst{1}
           \and L. Labadie
           \inst{6} 
           \and K. Perraut
           \inst{16}
           \and J. Sanchez-Bermudez
           \inst{1,18}
           \and
           C. -C. Lin\inst{1,19}
           \and A.~Amorim \inst{8,17} \and M.~Baub\"{o}ck \inst{9} \and 
           M.~Benisty\inst{16,11} \and
           J.P.~Berger \inst{16} \and  A.~Buron \inst{9} \and 
           P.~Caselli\inst{9}\and
           Y.~Cl\'enet \inst{10} \and  V.~Coud\'e~du~Foresto\inst{10}
           \and P.T. de Zeeuw\inst{15,9} 
           \and G. Duvert \inst{16}\and  W.~de~Wit\inst{11} \and  A.~Eckart\inst{6,12} \and  F.~Eisenhauer\inst{9} \and  M.~Filho\inst{7,8,11} \and  F.~Gao \inst{9}  \and  E.~Gendron\inst{10} \and  R.~Genzel \inst{9,13} \and   S.~Gillessen\inst{9} \and 
           R. Grellmann\inst{6} \and
           M. Habibi\inst{9} \and  X.~Haubois\inst{11} \and  F.~Haussmann\inst{9} \and 
           T.~Henning\inst{1} \and
           S.~Hippler\inst{1} \and  Z.~Hubert\inst{16} \and  M.~Horrobin\inst{6} \and  A. Jimenez Rosales \inst{9} \and  L.~Jocou\inst{16} \and  P.~Kervella\inst{10} \and  J.~Kolb\inst{11} \and  S.~Lacour\inst{10} \and  J.-B.~Le~Bouquin\inst{16} \and  P.~L\'ena\inst{10} \and 
           H. Linz\inst{1} \and
           T.~Ott\inst{9} \and  T.~Paumard\inst{10} \and  G.~Perrin\inst{10} \and  O.~Pfuhl\inst{4} \and 
           M. C. Ram\'{i}rez-Tannus\inst{1} \and
           C.~Rau\inst{9} \and  G.~Rousset\inst{10} \and  S.~Scheithauer\inst{1} \and  J.~Shangguan\inst{9} \and  J.~Stadler\inst{9} \and  O. Straub \inst{9} \and  C.~Straubmeier\inst{6} \and  E.~Sturm\inst{9} \and  E. van Dishoeck\inst{9,15} \and  F.~Vincent\inst{10} \and  S. von Fellenberg\inst{9} \and  F.~Widmann\inst{9} \and  E.~Wieprecht\inst{9} \and  M.~Wiest\inst{6} \and  E.~Wiezorrek\inst{9} \and  S.~Yazici\inst{9,6} \and G.~Zins\inst{11}
          }

%
\institute{ Max Planck Institute for Astronomy, K\"{o}nigstuhl 17, Heidelberg, Germany, D-69117 \\
  \email{maria.koutoulaki@eso.org}\\
  \and  Dublin Institute for Advanced Studies, 31 Fitzwilliam Place, D02\,XF86 Dublin, Ireland
   \and  School of Physics, University College Dublin, Belfield, Dublin 4, Ireland 
   \and  European Southern Observatory, Karl-Schwarzschild-Str. 2, 85748 Garching, Germany
   \and Department of Space, Earth \& Environment, Chalmers University of Technology, SE-412 93 Gothenburg, Sweden
  \and  I. Physikalisches Institut, Universität zu Köln, Zülpicher Str. 77, 50937, K\"{o}ln, Germany
  \and  Faculdade de Engenharia, Universidade do Porto, Rua Dr. Roberto Frias, P-4200-465 Porto, Portugal
  \and  CENTRA, Centro de Astrofísica e Gravitação, Instituto Superior Técnico, Avenida Rovisco Pais 1, P-1049 Lisboa, Portugal
  \and  Max Planck Institute for Extraterrestrial Physics, Giessenbachstrasse, 85741 Garching bei M\"{u}nchen, Germany
  \and  LESIA, Observatoire de Paris, Université PSL, CNRS, Sorbonne Université, Université de Paris, 5 place Jules Janssen, 92195 Meudon, France
  \and  European Southern Observatory, Casilla 19001, Santiago 19, Chile
  \and  Max-Planck-Institute for Radio Astronomy, Auf dem H\"{u}gel 69, 53121 Bonn, Germany
  \and  Department of Physics, Le Conte Hall, University of California, Berkeley, CA 94720, USA
  \and  Unidad Mixta Internacional Franco-Chilena de Astronomía (CNRS UMI 3386), Departamento de Astronomía, Universidad de Chile, Camino El Observatorio 33, Las Condes, Santiago, Chile
  \and  Sterrewacht Leiden, Leiden University, Postbus 9513, 2300 RA Leiden, The Netherlands
  \and  Univ. Grenoble Alpes, CNRS, IPAG, F-38000 Grenoble, France
  \and  Universidade de Lisboa - Faculdade de Ciências, Campo Grande, P-1749-016 Lisboa, Portugal
  \and  Instituto de Astronom\'{i}a, Universidad Nacional Aut\'{o}noma de M\'{e}xico, Apdo. Postal 70264, Ciudad de M\'{e}xico, 04510, M\'{e}xico 
  \and Institute of Astronomy, University of Hawaii, 2680 Woodlawn Drive, Honolulu, HI 96822, USA}
   \date{accepted in A\&A October 20, 2020}
%

\abstract
{51\,Oph is a Herbig Ae/Be star that exhibits strong near-infrared CO ro-vibrational emission at 2.3\,\mic, most likely originating in the innermost regions of a circumstellar disc.}
   {We aim to obtain the physical and geometrical properties of the system by spatially resolving the circumstellar environment of the inner gaseous disc.}
   {We used the second-generation Very Large Telescope Interferometer instrument GRAVITY to spatially resolve the continuum and the CO overtone emission. We obtained data over 12 baselines with the auxiliary telescopes and derive visibilities, and the differential and closure phases as a function of wavelength. We used a simple local thermal equilibrium ring model of the CO emission to reproduce the spectrum and CO line displacements.}
   %
  { Our interferometric data show that the star is marginally resolved at our spatial resolution, with a radius of $\sim 10.58\pm 2.65$\,\Rsun. The K-band continuum emission from the disc is inclined by 63\degr$\pm$1\degr, with a position angle of 116\degr$\pm$1\degr, and 4$\pm$0.8\,mas (0.5$\pm$0.1\,au) across. The visibilities increase within the CO line emission, indicating that the CO is emitted within the dust-sublimation radius. By modelling the CO bandhead spectrum, we derive that the CO is emitted from a hot (T=1900--2800\,K) and dense (\Nco=(0.9--9)$\times$10$^{21}$\,cm$^{-2}$) gas. 
   The analysis of the CO line displacement with respect to the continuum allows us to infer that the CO is emitted from a region 0.10$\pm$0.02\,au across, well within the dust-sublimation radius. The inclination and position angle of the CO line emitting region is consistent with that of the dusty disc.}
   %
   { Our spatially resolved interferometric observations confirm the CO ro-vibrational emission within the dust-free region of the inner disc. Conventional disc models exclude the presence of CO in the dust-depleted regions of Herbig AeBe stars. Ad hoc models of the innermost disc regions, that can compute the properties of the dust-free inner disc, are therefore required.}
   

\maketitle

%
\section{Introduction} 

Circumstellar discs are crucial for understanding how stars and planets form. 
In spite of the enormous wealth of data provided by a new generation of instruments, such as Atacama Large Millimeter/submillimeter Array (ALMA) and VLT Spectro-Polarimetric High-contrast Exoplanet REsearch (VLT/SPHERE), 
understanding disc properties and evolution is still challenging. This is mostly due to the lack of knowledge about the physical properties and structure of the innermost disc regions. Even if these new instruments would allow us to spatially resolve the disc emission in the dust continuum and in a number of molecular lines down to scales of a few au, it is still demanding to resolve the structure of the innermost disc regions, and this has only been possible for a few objects \citep{Kraus2008, Eisner2007, Lazareff2017, Perraut2019}.  
%
%
%
Still, this is a region of great importance, as we expect that both the accretion of disc matter onto the central star and the ejection of magnetically controlled winds occur in a region smaller than about one au, even for young stars in nearby star-forming regions (~120-140 au). Spatially resolving the inner disc requires optical interferometry. The accessible tracers are limited to the continuum emission of dust close to the sublimation radius and to a few emission lines, the most prominent of which is the HI\,Br$\gamma$ line. It has been found so far that with very few exceptions, the Br$\gamma$ line mainly originates in a wind \citep{Weigelt2011,Alessio2015, Rebeca2015}. In a few objects, however, the overtone ro-vibrational emission of CO at $\sim$2.3\,\mic\ is detected and has been proved to  trace gas in Keplerian rotation very close to the star,  very likely  in the disc itself \citep{Carr1989,Kraus2000,Bik_Thi2004,Tatulli2008,CarattioGaratti2020}.

By model-fitting the line profile of the CO bandhead emission obtained with high spectral resolution observations, several authors showed that the emission comes from a relatively warm ($T\sim$2000-5000\,K) and dense gas (N$_{\mathrm{CO}}\sim10^{20}-10^{22}$cm$^{-2}$) in Keplerian rotation \citep{Carr1989,Najita1996,Ilee2014,Koutoulaki2019}. 
However, this information has so far not been used to further constrain the physical, chemical, and thermal processes occurring in the inner disc. This is mostly because we lack of a solid knowledge of the stellar properties and disc geometry, which prevents an accurate estimation of the location of the CO emission from the inferred Keplerian velocity.
%
Further progress requires directly measuring the location of the CO emission, using near-IR interferometry.


In this paper, we present a near-IR interferometric study of the star
51\,Oph (HD\,158643) using the ESO VLTI instrument GRAVITY. 51\,Oph  is a fast-rotating star  \citep[\vsini= 267$\pm$5 km/s][]{Dunkin1997} of about 4 \Msun, located at a distance D=123$\pm$4\,pc \citep{GAIA2018}. The spectral type is  B9.5-A0 \citep{Dunkin1997, Gray1998} and the luminosity is $\sim$ 250 \Lsun \citep{VandenAncker2001}. 
The spectrum is rich in molecular and atomic emission lines in the mid- and far-IR \citep{Thi2013_51Oph}.  
51 Oph shows bright 2.3 \mic\ CO overtone emission \citep{Thi2005,Berthoud2007,Tatulli2008}. \citet{Tatulli2008} spatially resolved the CO overtone emission in this star, using the first generation of near-IR interferometric instrument, VLTI/AMBER. They find that the CO emission originates within 0.15 au of the star and rotates in a Keplerian manner. 
However, the number of baselines, the resolution, and the signal-to-noise ratio of these early near-IR interferometric studies were limited, and therefore failed to achieve well-constrained physical properties of the disc-emitting region. 

The structure of the paper is as follows. 
The observations and data reduction are discussed in Section 2, and the results, modelling, discussion, and conclusions are presented in Sections 3, 4, 5 and 6 respectively.

\section{Observations and data reduction}
\label{sec:obs}
51 Oph was observed using the VLTI with the K-band beam combiner \textit{GRAVITY}  \citep{GRAVITY2017} on two dates, 29 May 2017 and 15 August 2017. The observations were performed using the 1.8\,m auxiliary telescopes (ATs) with the configurations B2-K0-D0-J3 and A0-G1-J2-K0 in May and August, respectively. The resulting projected baselines and position angles (PAs) are shown in Table\,\ref{tab:obs}.
The observations were performed in single-field high-combined mode of GRAVITY, in which the visibilities of the science target are recorded on both the fringe-tracker detector (at low spectral resolution, $\mathcal{R}=30$) and the science detector (at high spectral resolution, $\mathcal{R}=4000$). The detector integration time (DIT) on the fringe tracker and science detector was set to 0.85\,ms and 30\,ms, respectively, with a total exposure timne of 5\,min per block on the science detector. This resulted in a total integration time on source of 2\,hr and 1\,hr for the first and second night, respectively. 

The data were reduced using the instrument pipeline (version 1.0.11). In order to calibrate the transfer function, the calibrator star HD\,163955 was used. The wavelength calibration was refined using the atmospheric telluric absorption lines present in the spectra of both target and calibrator. A shift of 7\,\AA\ was found and taken into account. In addition, the spectrum of the calibrator was used to correct the target spectrum for the absorption telluric features and the instrumental response. Finally, the 2MASS K-band magnitude of K=4.3 was adopted to flux calibrate the spectrum of 51 Oph.

For the continuum analysis, the fringe-tracker data were used, as the fringe tracker operates at higher frequencies, making it less sensitive to the atmospheric changes. There is no significant difference between the level of the calibrated visibilities measured by the fringe-tracker and the science detector.

\begin{table*}
\caption{Log of the VLTI/\textit{GRAVITY} observations}             
\label{tab:obs}      
\centering     
\scalebox{0.94}{
\begin{tabular}{c c c c c c c c c c}     
\hline\hline       
Object & Observation& 
 AT$\rm^{a}$ &Baseline& PA& Spectral&Wavelength &  UD\\ 
Name & Date &
Array &Range&& Mode$\rm^{b}$&  & Diameter$\rm^{c}$\\
&&&(m)&(\rm$^{o}$)&&($\mu$m)&(mas) \\
\hline                    
  51\,Oph&2017 May. 29&B2-K0-D0-J3&102/56/135/91/34/117&46/163/43/79/34/67&HR-K&1.99-2.47& \\
  HD\,163955&2017 May. 29&B2-K0-D0-J3&102/56/135/91/34/117&46/163/43/79/34/67&HR-K&1.99-2.47&0.38 \\
  51 Oph&2017 Aug. 15&A0-G1-J2-K0&49/87/110/52/111/81&6/37/84/66/109/135&HR-K&1.99-2.47& \\
  HD\,163955&2017 Aug. 15&A0-G1-J2-K0&49/87/110/52/111/81&6/37/84/66/109/135&HR-K&1.99-2.47&0.38 \\
  
\hline 
\end{tabular}}
\tablefoot{
$^{a}$ Auxiliary Telescope, 
$^{b}$ High spectral resolution in K-band, 
$^{c}$ Uniform-disc diameter derived from the software package SearchCal from Jean-Marie Mariotti Center (JMMC). }
\end{table*}

\section{Results}
\begin{figure*}
     \centering
     \includegraphics[width=\linewidth]{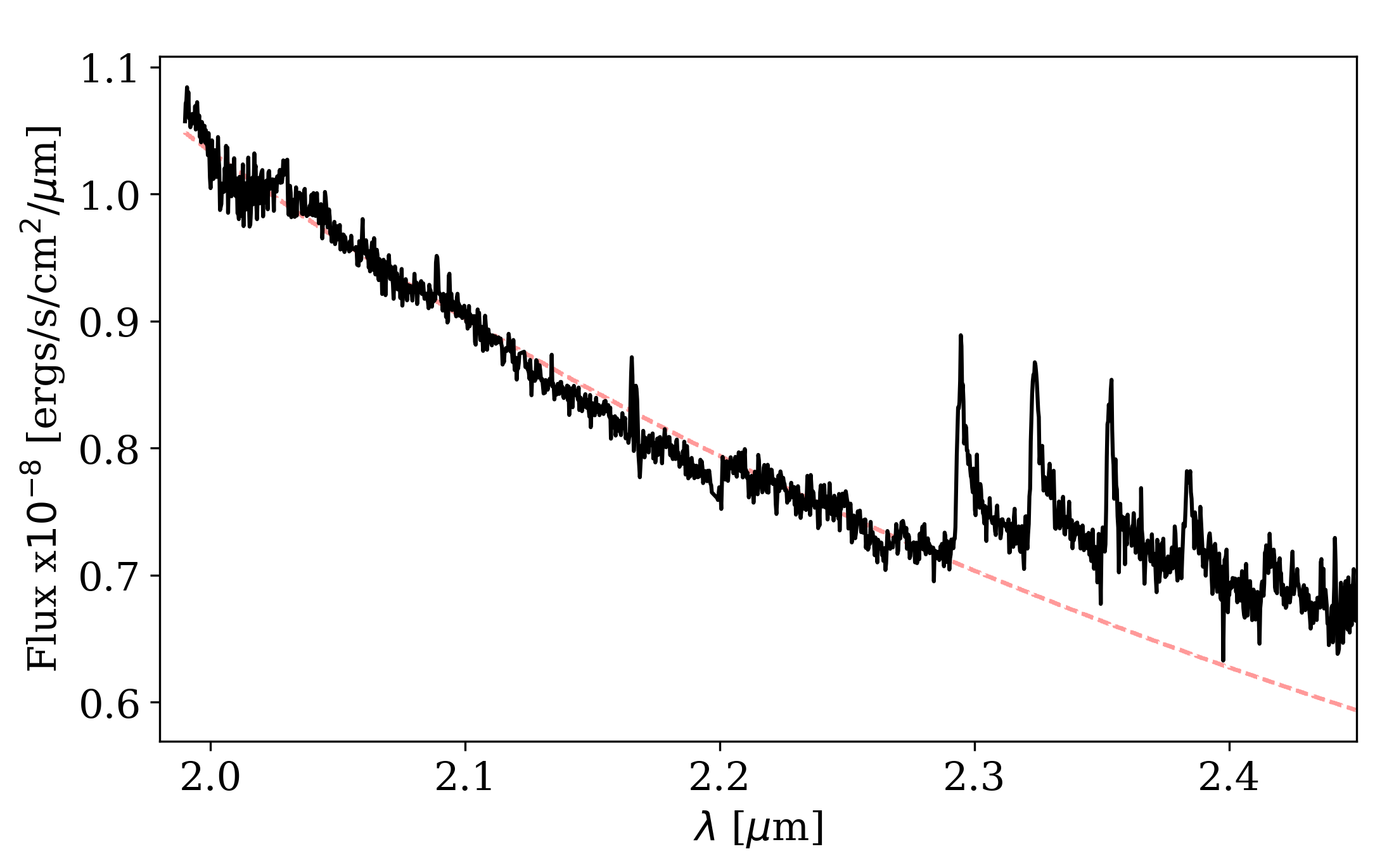}
     \caption{K-band flux-calibrated spectrum of 51\,Oph for the May 2017 dataset. The dashed red line shows the adopted continuum distribution, computed as the sum of a black body at $T=10000$ K, which contributes 80\% to the total flux, and a 20\% contribution from a black body at $T=1500 K$, approximately the silicate sublimation temperature \citep{Lazareff2017, Perraut2019}.}
     \label{fig:full_spec_51Oph}
 \end{figure*}

For each epoch, GRAVITY observations of 51\,Oph provide us with a spectrum, six spectrally dispersed visibilities and differential phases, and four closure phases  (Fig.\,\ref{fig:May+Aug_int}). In both epochs, the K-band spectrum clearly shows the CO overtone emission and  bright \HI\,\brg. Figure~\ref{fig:full_spec_51Oph} shows the  flux-calibrated spectrum of 51 Oph for the May dataset. The continuum K-band contribution is also overplotted (dashed red line) in the same figure. This continuum was used to derive the continuum-subtracted spectra in the text (for the May and August spectra) and was computed as the sum of a black body emission at the effective temperature of the star (T=10\,000\,K, \citealt{Gray1998}), which contributes 80\% to the total K-band flux, plus additional black body emission at a temperature of 1500\,K, representing the K-band dust emission, and contributing 20\% to the total K-band flux \citep[][hereafter Paper I]{Lazareff2017, Perraut2019}. No sign of variability is observed between epochs. However, the August dataset is about three times noisier than the May dataset (see Fig.\,\ref{fig:CO_spectrum}).We focus this paper on the CO emission, and defer the analysis of \brg\ to a following paper. 
%

\subsection{CO spectrum}

\begin{figure*}
    \centering
    \includegraphics[width=1.0\textwidth]{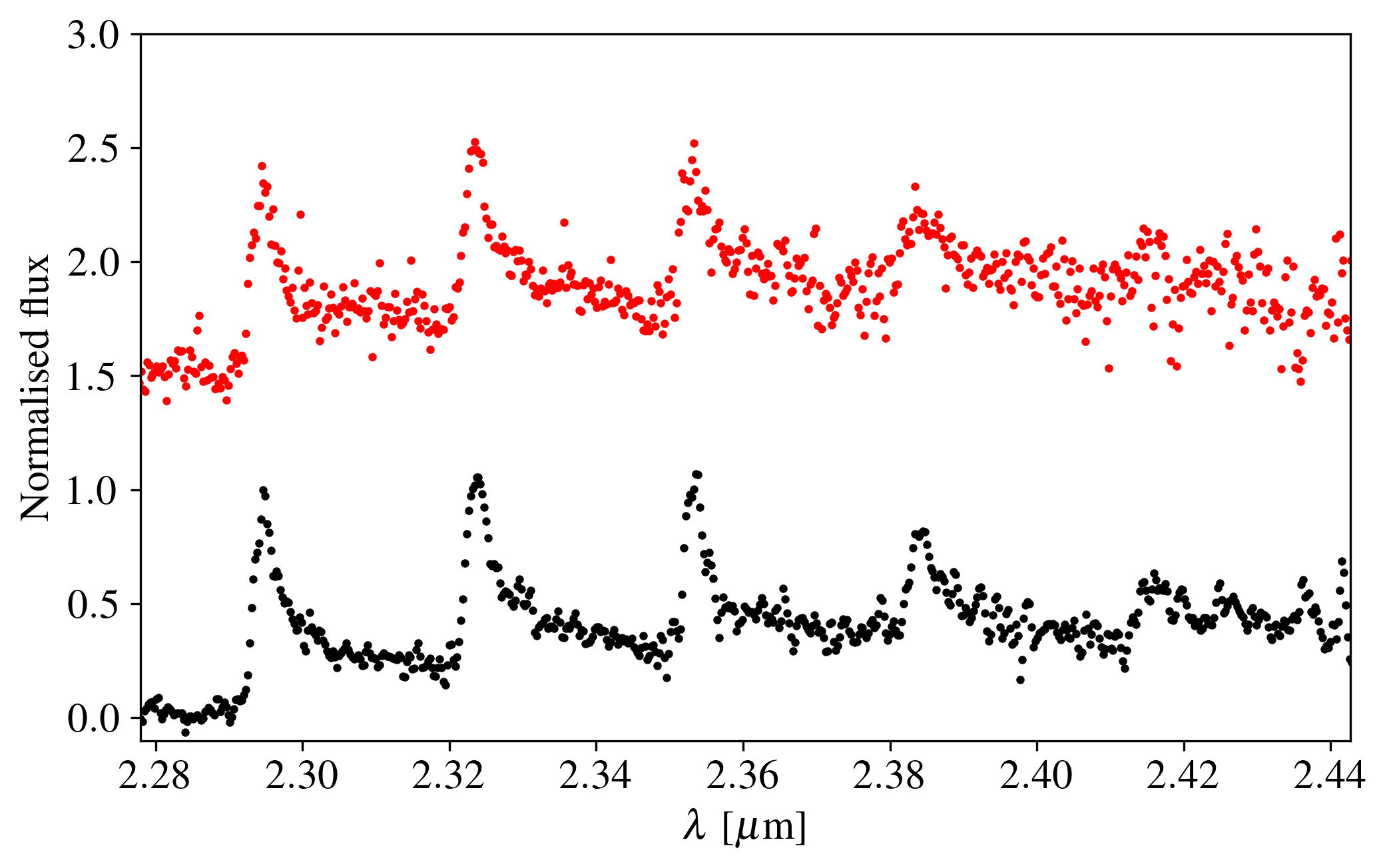}
    \caption{ May 2017 CO spectrum of 51\,Oph (black) and August 2017 (red) showing four ro-vibrational transitions ($\upsilon\mathrm{=2-0, 3-1, 4-2,\,and\, 5-3}$) and an indication of the fifth ($\upsilon\mathrm{=6-4}$). The spectra are continuum-subtracted and normalised to the peak of the first bandhead. The August data are shifted by 1.5 for comparison.}
    \label{fig:CO_spectrum}
\end{figure*}

 The 51 Oph continuum-subtracted spectrum is shown in Fig.\,\ref{fig:CO_spectrum} for the May (black) and August (red) 2017 datasets. The emission of the first four ro-vibrational transitions ($\upsilon\mathrm{=2-0, 3-1, 4-2,\, and\, 5-3}$), and an indication of the fifth ($\upsilon\mathrm{=6-4}$) is clearly observed.

The profile of the first three  bandheads shows a sharp peak at similar intensity, while the last two have slightly lower peak values. The so-called blue shoulder is observed in the first CO bandhead even at the low spectral resolution of GRAVITY. The blue shoulder is very likely due to Keplerian broadening \citep[e.g.][]{Chandler1995}, which causes a broadening towards shorter wavelengths of the otherwise intrinsic sharp flux increase of the bandhead. 
%
No individual or blended CO J components are observed in the band tails because the spectral resolution and signal-to-noise-ratio of our spectra are not high enough, and also because the many telluric atmospheric absorption lines in this region of the spectrum were imperfectly removed.  
As previously mentioned, 
the spectra at the two epochs are very similar, with marginal differences in the third and fourth bandheads that are well within the uncertainties.   


\subsection{Interferometric observables}

\begin{figure*}[ht]
\includegraphics[width=\textwidth]{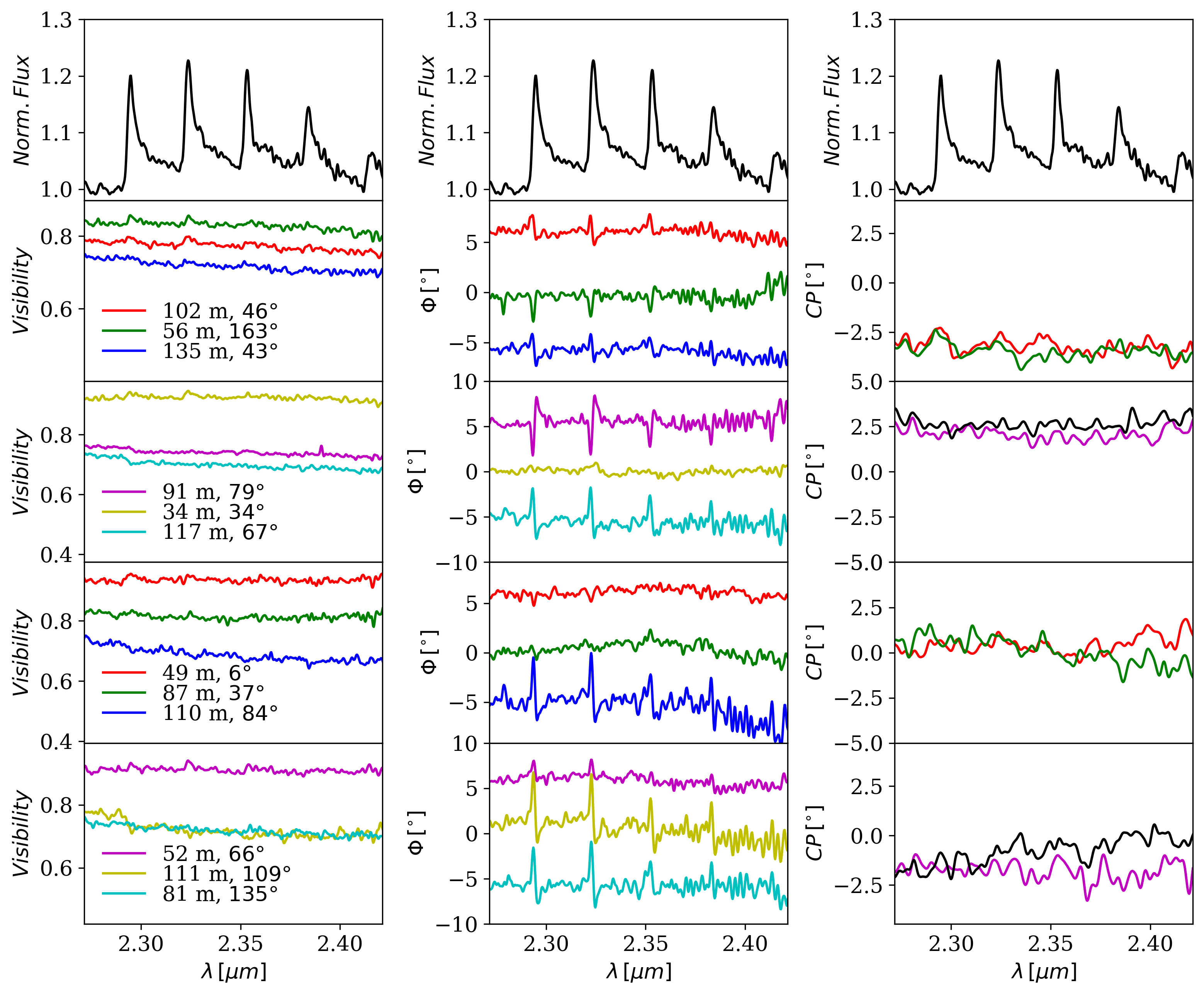}
\caption{GRAVITY interferometric data for May and August 2017. The top panels show the CO spectrum  of the May dataset, normalised to the  continuum. Visibilities, differential phases, and closure phases are displayed in the left, middle, and right panels, respectively. The differential phases are shifted by $\pm 5$\degr for clarity and are in units of degrees. Closure phases are in units of degrees. The data are smoothed to a resolution of R=2000.}
\label{fig:May+Aug_int}
\end{figure*}

Visibilities, differential and closure phases were measured in 51\,Oph at 12 different baselines (6 for each epoch, see Fig.\,\ref{fig:May+Aug_int}). 

The visibilities in all the baselines show that the continuum is marginally resolved with values ranging from 0.7 to 0.9. 
In several baselines, the visibility at the peak of the bandheads  is slightly higher than that of the continuum, suggesting a smaller emitting region. 

A clear differential phase signal is detected in the first  four CO bandheads for most  baselines in both epochs, indicating a clear shift of the photo-centre of the CO emission with respect to that of the continuum. 

No closure phase is detected within the calibration error ($\sim5^\circ{}$), indicating a symmetric circumstellar environment.

\subsection{Circumstellar environment: The continuum emission}

 The observed continuum emission is the sum of the emission of the central star and that of surrounding circumstellar matter, which very likely originates from a disc. The relative flux and size of the two components are different, and the resulting observed visibility $V_c$ can be written as

\begin{equation}
    V_{c} = \frac{F_{*} V_{*}+ F_\mathrm{{disc}}V_\mathrm{{disc}}}{F_{c}}
\end{equation}
where $F_{*}$ and $V_{*}$ are the flux and visibility of the star, $F_\mathrm{{disc}}$ and $V_\mathrm{{disc}}$ are  the flux and visibility of the disc, and  $F_{c}=F_{*}+F_\mathrm{{disc}}$ is the total continuum flux. In 51 Oph, the dominant flux contribution in the K band is that of the star, which provides  80\% of the total flux (see Sec.3).
The total continuum visibility (star plus disc) is  fitted with a two-Gaussian model: a one-dimensional Gaussian for the star, with the stellar angular radius as the single free parameter; and a two-dimensional Gaussian for the disc, characterised by three free parameters:  
 radius,  inclination, and position angle of the major axis of the disc.

\begin{figure*}
    \centering
    \includegraphics[width=1.0\columnwidth]{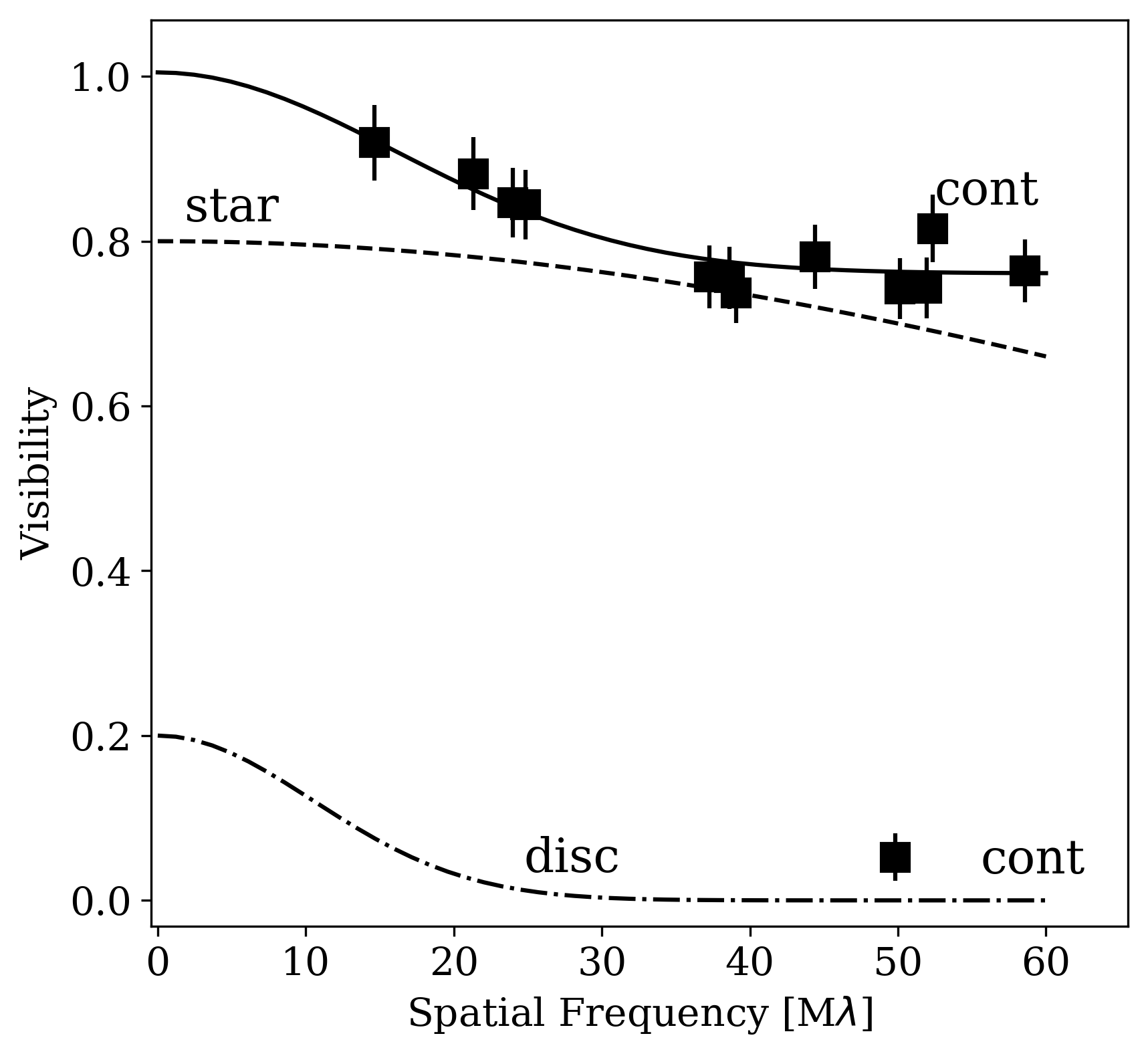}
    \includegraphics[width=1.0\columnwidth]{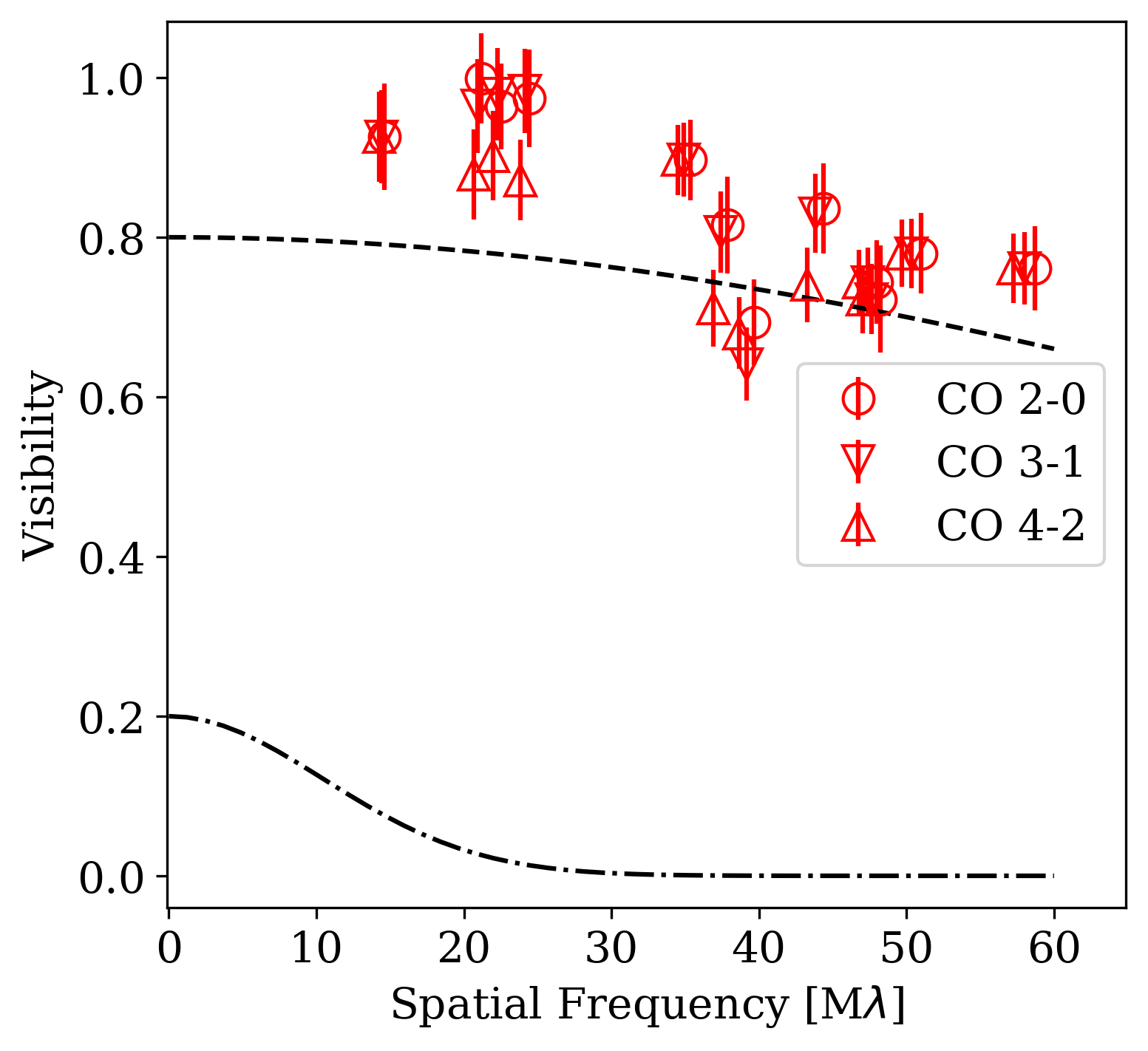}
    \caption{\textit{Left:} Visibility vs. spatial frequency  for the continuum (black squares). 
    The best fit to  the total continuum emission (star+disc) is  shown by the solid black line,  the  contribution of the star by the dashed line, and that of the disc by the dash-dotted line. \textit{Right:} For comparison, we show  the visibility at the peak of the 2-0 CO bandhead (filled red circles), the 3-1 CO bandhead (open red circles), and the 4-2 CO bandhead (red triangles). The black lines are the same as in the left panel.}
    \label{fig:gauss_fit}
\end{figure*}

Our best-fitting model gives a stellar radius (half width at half maximum) of $R_*$=$0.4\pm 0.1$\,mas and a disc radius of $R_\mathrm{disc}$=$4.0\pm0.8$\,mas with an inclination ($i$) and major-axis PA of $i$=52\degr$\pm$1\degr\ and PA=116\degr$\pm$1\degr, respectively  (see Table\,\ref{tab:cont_fit} and  Fig.\,\ref{fig:gauss_fit}, solid black line). At the distance of 51\,Oph, this translates into a stellar radius of $R_*$=0.05$\pm$0.01\,au, or $R_*$=10.6$\pm$2.6\,\Rsun, and a disc emission about ten times higher (i.e. $R_{\mathrm{disc}}$=0.49$\pm$0.10\,au).

Our measurement of the stellar radius is in agreement with the  measurements obtained at 6000 \AA with  the CHARA interferometer by \cite{Jamialahmadi2015}, 
although the CHARA observations, thanks to the much longer baselines, reveal a highly  flattened stellar photosphere of radius 0.42$\pm$0.01 x 0.6$\pm$0.05\,mas.
It should be mentioned that, as discussed in \cite{Jamialahmadi2015}, the stellar luminosity  computed from the interferometric size and the effective temperature derived from the spectral type of 51 Oph would be about $10^3$ \Lsun, four times higher than the values  derived from the observed  SED \citep{VandenAncker2001}.
This  suggests that 51\,Oph is a giant star rotating close to break-up, either in the pre-main sequence and still in the contraction phase, or a more evolved star already in the post-main sequence evolutionary phase (see \citealt{Jamialahmadi2015} for more details). 

Our derived disc radius, inclination, and PA are also roughly in agreement with those found in Paper I from the same set of data using a variety of different geometrical models to account for the disc emission. The main difference between our values and those found in Paper I comes from the fact that in the latter, the star was assumed to be a point source (i.e. $V_*$=1). 

As discussed  in detail in Paper I, the  $R_{\mathrm{disc}}$ value is consistent within the uncertainties with the expected location of the silicate sublimation region. When a stellar luminosity of 250\,L$_\odot$ and a silicate sublimation temperature of 1500\,K are assumed, the dust sublimation radius is located between 0.5\,au to 1.7\,au, depending on whether small (cooling efficiency $\epsilon$=0.1) or large (cooling efficiency of $\epsilon\sim$1) silicate grains are considered \citep[see Paper I,][ eq.\ $R_{subl}=\sqrt{\frac{L^{*}}{16\pi \sigma \epsilon T^4}}$]{Kama2009, Isella2005, Dullemond2001}.


%
\begin{table}[ht]
\caption{Properties of the continuum-emitting region}  
\vspace{1.0 cm}
\label{tab:cont_fit}      
\centering                          
\begin{tabular}{l c c}        
\hline\hline
\noalign{\smallskip}
Parameter & \multicolumn{2}{c}{Value} \\
\hline
\noalign{\smallskip}
R$_{*}$& 0.4$\pm$0.1 mas &10.6$\pm$2.6\,\Rsun\\
R$_\mathrm{{disc}}$& 4.0$\pm$0.8 mas&0.49$\pm$0.10 au\\
$i$& 63 $^{\circ}\pm 1^{\circ a}$ \\
PA & 116$^{\circ}\pm 1^{\circ a}$ \\
   \hline                                   
\end{tabular}
\tablefoot{$^{a}$ from Paper I}
\end{table}

\subsection{Circumstellar environment: The CO emitting region}

To compute the size of the CO line emitting region, we subtracted the contribution of the star and the disc and defined the continuum-subtracted visibility at any given wavelength within the CO bandheads $V_\mathrm{{line}}$ as 
\begin{equation}
    V_\mathrm{{line}} = \frac{\sqrt{|V_\mathrm{{tot}}F_\mathrm{{tot}}|^2+|V_{c}F_{c}|^2-2V_\mathrm{{tot}}F_\mathrm{{tot}}V_{c}F_{c}\cos{\Phi}}}{F_\mathrm{{line}}}
    \label{eq:pvsi_phi}
\end{equation}
In this expression, $F_\mathrm{{line}}$ is the line flux,  $F_\mathrm{{tot}}=F_{c}+F_\mathrm{{line}}$ is the observed flux, $V_\mathrm{{tot}}$ is the observed visibility,  $V_c$ and $F_c$ are the continuum visibility and the continuum flux, and $\Phi$ is the observed differential phase signal \citep[see][for more details]{Weigelt2011}. The continuum visibility is measured in the nearest line-free wavelength interval, and the continuum flux is taken at each wavelength from the continuum fit shown in Fig. \ref{fig:full_spec_51Oph} (dashed red line). $V_\mathrm{{line}}$ and $F_\mathrm{{line}}$ depend on the observed continuum properties (visibility and line-to-continuum ratio), not on the continuum deconvolution in stellar + disc components.
In our observations, the differential phase is very small ($<$5\degr). Thus E.q. \ref{eq:pvsi_phi} can be approximated as
\begin{equation}
    F_\mathrm{{tot}}V_{\mathrm{tot}} = F_{c}V_{c}+F_\mathrm{{line}}V_\mathrm{{line}}
    \label{eq:pvis}
\end{equation}
The errors are calculated by propagating Eq. \ref{eq:pvis}.

The continuum-subtracted visibilities at the peak of the first and second bandheads are computed by averaging over three consecutive frequencies at each bandhead. The results are shown in  Fig.\,\ref{fig:gauss_fit}. The continuum-subtracted visibilities for the first and second bandheads are very similar, and in all cases, they are slightly higher than the continuum visibilities. A 2D elliptical Gaussian fitting to the continuum-subtracted CO visibilities (assuming the inclination and PA of the disc) gives a radius of the CO emitting region of $R_\mathrm{{CO}}$=0.7\,mas (i.e. 0.09 au at a distance of 123\,pc). This value is much lower than the continuum disc radius. However, given that the CO is barely resolved,  it  should be considered as a rough estimate of the real size of the CO emitting region. 
 
 
 A much more accurate estimate of the geometrical properties of the CO emitting region can be derived from the differential phase data (Fig.\,\ref{fig:May+Aug_int}, $middle panel$). As for the visibility, the continuum contribution can be removed from the observed values.
 The continuum-subtracted  differential phase ($\Delta \phi$) at any wavelength within the line is computed  following the expression \citep{Weigelt2007}
\begin{equation}
\sin(\Delta\Phi)=\sin(\Phi)\cdot \dfrac{|F_\mathrm{{tot}}V_\mathrm{{tot}}|}{|F_\mathrm{{line}}V_\mathrm{{line}}|},
\end{equation}
with the same symbols as in Eq.2. 
The displacement of the photo-centre of the emission at any given wavelength ($\delta$) is then computed as:
\begin{equation}
\delta = -\Delta \Phi \frac{\lambda}{2 \pi B},
\end{equation}
where $B$ is the length of the baseline.

\begin{figure*}
    \centering
    \includegraphics[width=1.0\textwidth]{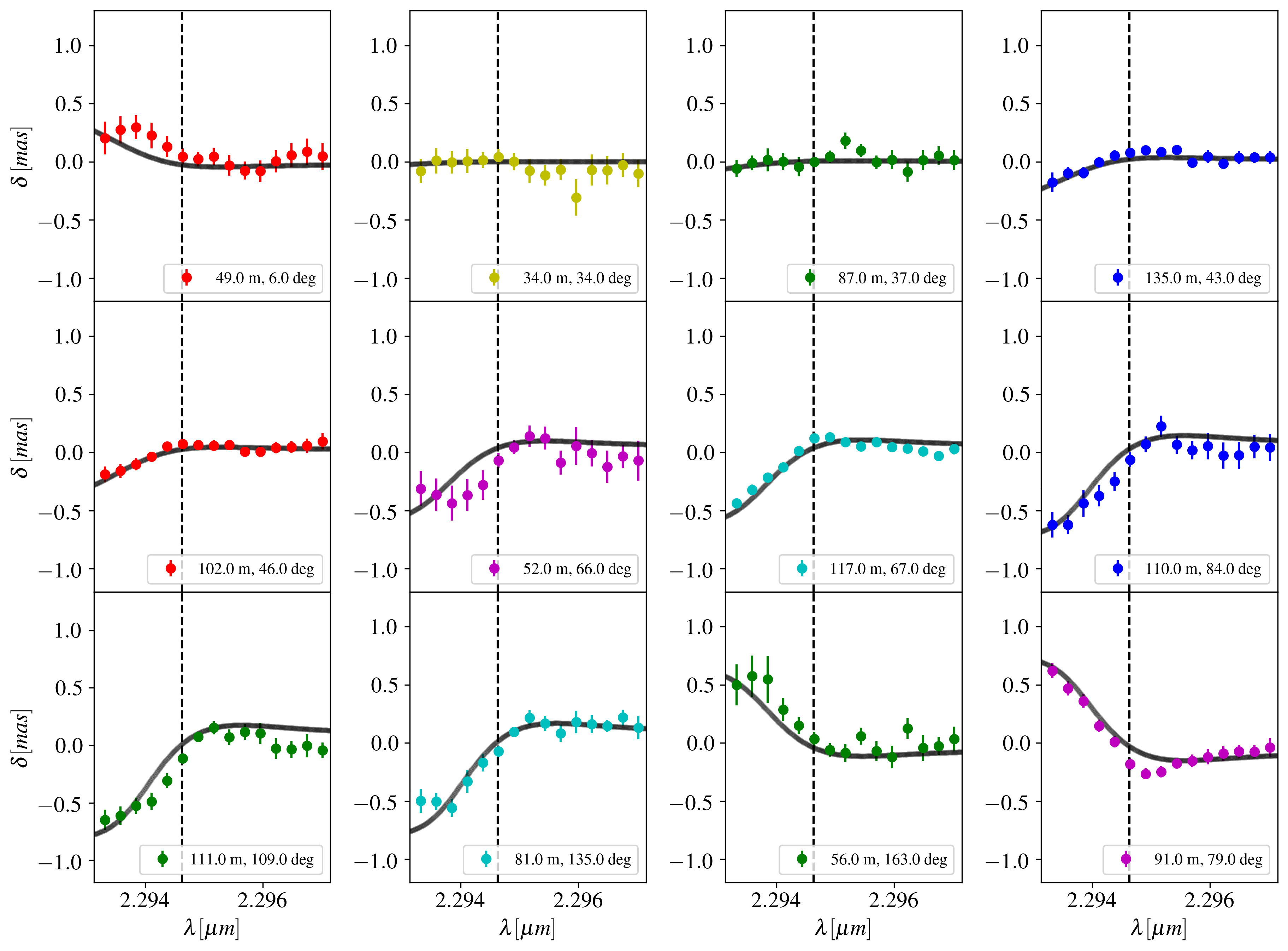}
    \caption{Comparison of the observed wavelength-dependent displacement $\delta$ (filled circles) and best model predictions (thick black lines) for the first CO bandhead. Each panel refers to a different projected baseline, as labelled. At the distance of 51 Oph (123 pc) 0.8\,mas corresponds to 0.1\,au. The shifts are calculated for a ring of radius 0.1\,au, inclination of 70\degr, and $v_{rot}$=140\,\kms. The vertical dashed line shows the peak of the CO bandhead. The physical parameters of the model are T=2400\,K, \Nco=$4\times10^{21}$\,cm$^{-2}$, \deltav=15\,\kms, and \vsini= 130\,\kms.}
    \label{fig:pshifts_model_obs_1st}
\end{figure*}

\begin{figure*}
    \centering
    \includegraphics[width=1.0\textwidth]{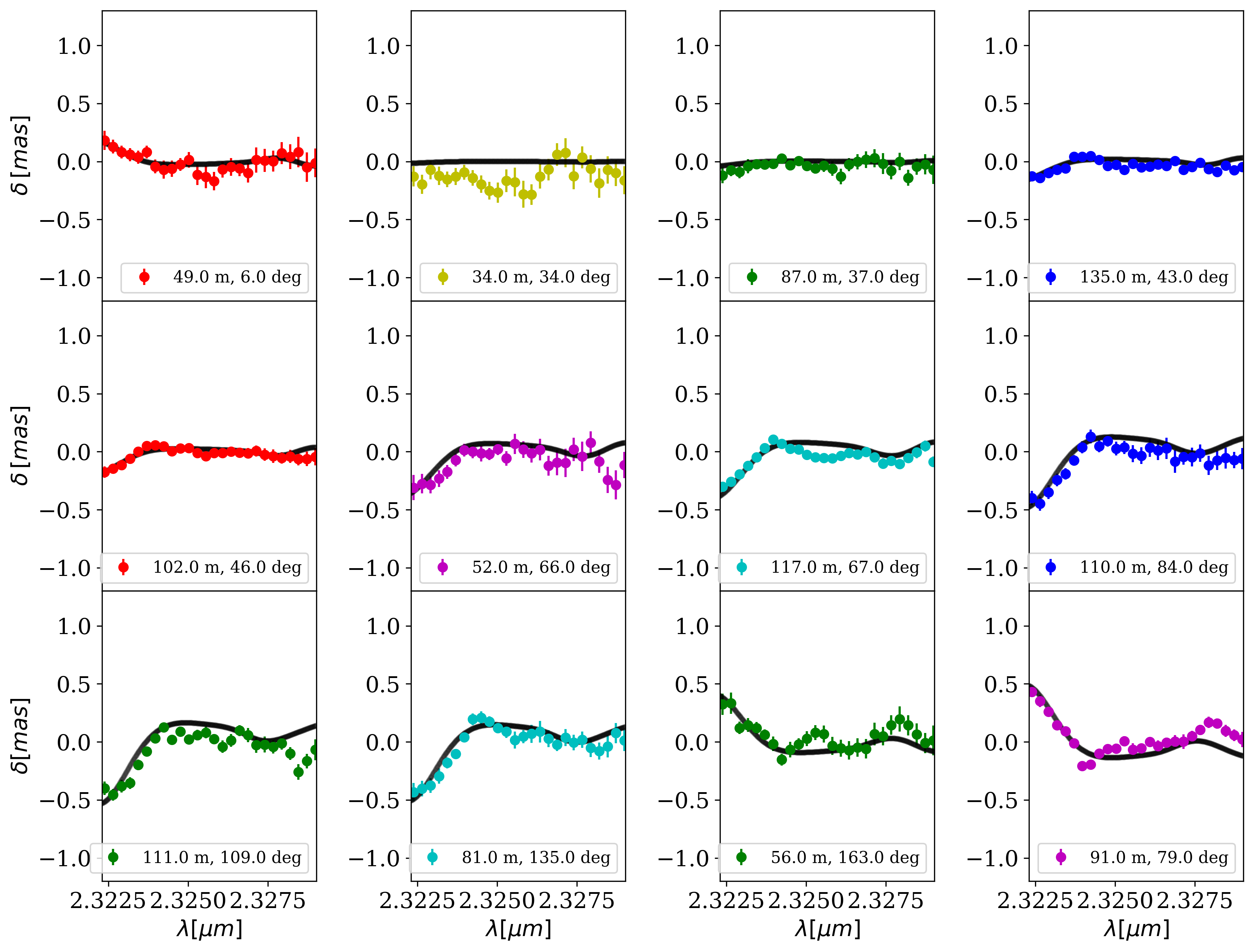}
    \caption{ Same as Fig.\,\ref{fig:pshifts_model_obs_1st}, but for the second bandhead.}
    \label{fig:pshifts_model_obs_2nd}
    \end{figure*}

\begin{figure*}
    \centering
    \includegraphics[width=1.0\textwidth]{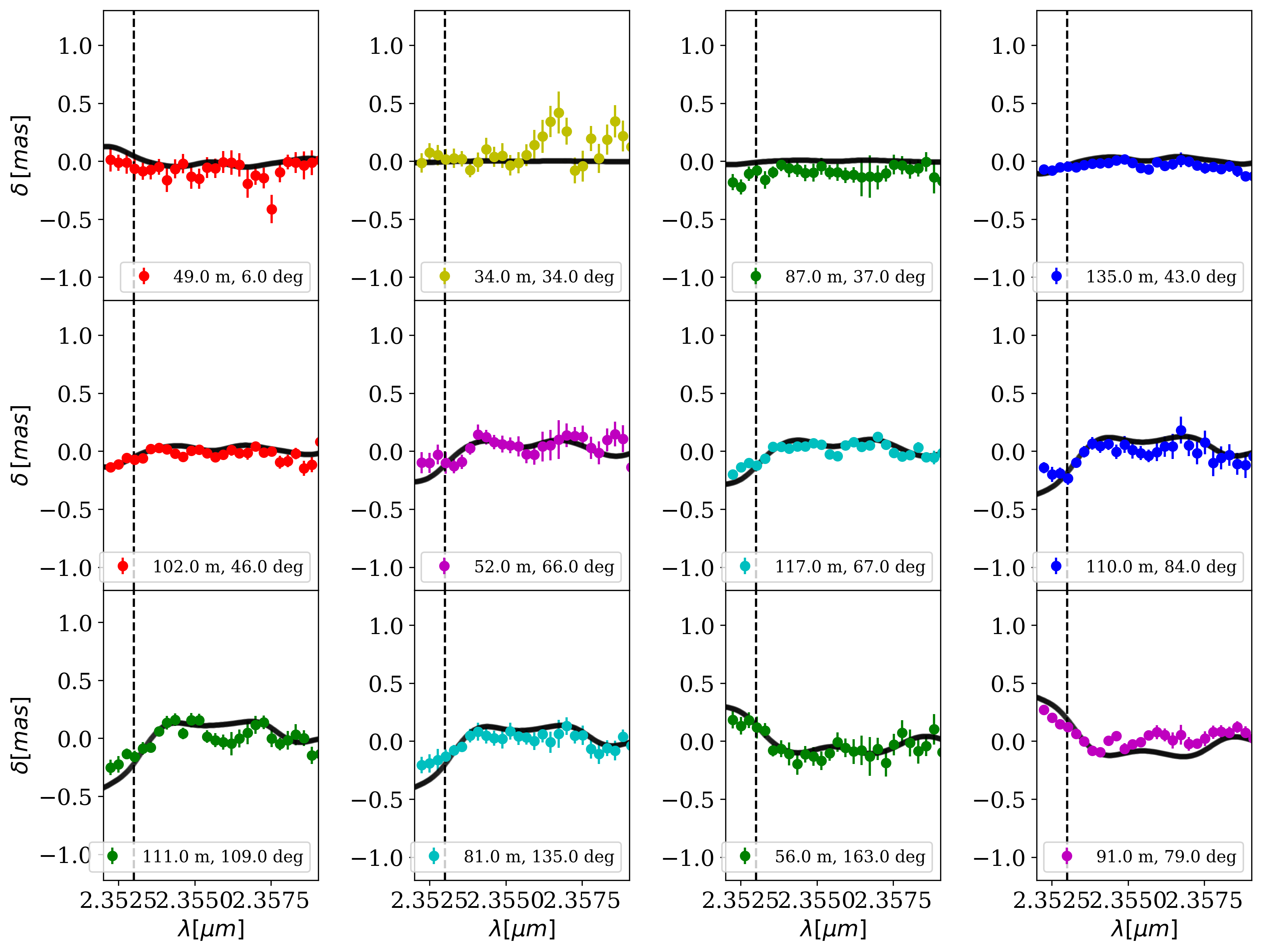}
    \caption{ Same as Fig.\,\ref{fig:pshifts_model_obs_1st}, but for the third bandhead.}
    \label{fig:pshifts_model_obs_3rd}
\end{figure*}

The $\delta$ values computed for the first three bandheads as a function of the wavelength are shown in Fig.\,\ref{fig:pshifts_model_obs_1st}, \ref{fig:pshifts_model_obs_2nd}, and \ref{fig:pshifts_model_obs_3rd} for the 12 observed projected baselines (PBLs).
Under the assumption that the continuum emission is centrally symmetric, the quantity $\delta$ provides a measurement of the photocentre displacement of the line-emitting region at the specific wavelength within the line with respect to the continuum. In the simplest case of an individual line, such as HI\,\brg\,, emitted by a tilted rotating ring, $\delta$ will vary from a maximum value in the blue wing of the line to a minimum in the centre and again to a maximum of opposite sign in the red wing. This typical behaviour has been used to measure the size, inclination, and PA of the \brg\ emitting region, as well as the rotation direction, in the Herbig Ae HD163296 \citep[e.g.,][]{Ellerbroek2015}.  The case of the CO bandheads is more complex, as at any given wavelength the emission is due to the overlap of various blue- and red-shifted individual $J$ components. As a consequence, the simple association of wavelength and location of the emitting region is lost  \citep[see e.g.][for more details]{Koutoulaki2019} and a proper modelling of the bandhead emission is required  to interpret the observed displacement of the CO emission with wavelength (see Sect.\,4.2). 


\section{Modelling}

\subsection{Modelling the CO spectrum}
\label{sec:COmodel}

Following \cite{Koutoulaki2019}, and \cite{Kraus2000}, we computed synthetic spectra and intensity maps of the CO bandhead emission as a function of wavelength. 

In brief, the CO emitting region was modelled as a narrow ring of gas with temperature ($\textit{T}$), and column density ($N_{\rm CO}$),  rotating at a Keplerian velocity $v_{rot}$, around the central star. In each vibrational band, we considered 100 rotational transitions; each \textit{J} component has a Gaussian profile of width $\Delta \rm v$ (the so-called intrinsic line width). 
We created a spectrum over the range covered by the observed bandheads, with very high spectral resolution ($\mathcal{R}=100000$), so that each individual \textit{J} transition profile is well sampled. We first computed  the total optical depth at each wavelength, then  the intensity  given by $I(\nu, \textit{J})=B_{\nu}(T)(1-e^{-\tau_{\nu,\textit{J}}})$. 
The resulting spectrum was then convolved over the Keplerian velocity pattern, reduced to the instrumental spectral resolution of $\mathcal{R}=4000$, and normalised to the peak of the first bandhead. The bandhead profiles depend on four free parameters: the temperature ($\textit{T}$), the  column density of CO ($N_{\rm CO}$), the intrinsic line width $\Delta \rm v$, and the Keplerian velocity of the ring projected along the line of sight ($v_{rot} \sin {i}$).

Even at the GRAVITY spectral resolution of $\mathcal{R}=4000$, the shape of the bandheads provide some interesting constraints to the physical properties and location of the CO emission. To illustrate this, we computed spectra by varying the four free parameters over a wide range of values, and we estimated the quality of the fit by considering how well different specific spectral features are reproduced. A series of examples on how the CO bandhead spectrum changes as function of the different free parameters, and the corresponding comparison with the observed spectrum is shown and discussed in Appendix A. Our main findings are described below.

 The first three bandheads observed in 51\,Oph are due to the overlap of very optically thick individual J components (see Fig.\,\ref{fig:tau_plot}). As a consequence, the two parameters $\textit {T}$ and $N_{\rm CO}$ have large uncertainties because an increase in $T$ can be compensated for by a decrease in  $N_{\rm CO}$, for instance (see Figs.\,\ref{fig:spec_err_T} and \ref{fig:spec_err_NCO}). The fourth and fifth bandheads are optically thinner than the first three bandheads and therefore can give better constraints on $\textit {T}$ and $N_{\rm CO}$ (see Fig.\,\ref{fig:tau_plot}). However, the line profiles of these two bandheads are more affected by the poor atmospheric transmission at the edge of the K band which makes their line profiles noisier and thus less reliable. We find that the observed spectrum is well reproduced by a range of models with pairs of temperature and column density values ranging from T= 1900--2800\,K and $N_{\rm CO}$= 9$\times 10^{20}$--9$\times 10^{21}$ (see Fig.~\ref{fig:spec_fit_ddif_T_NCO}). The range of acceptable pairs of $\textit {T}$ and $N_{\rm CO}$ was selected by simultaneously increasing (decreasing) the $\textit {T}$ and decreasing (increasing) the $N_{\rm CO}$ values until the observed spectrum (peak and/or tail of the bandheads) differed by more than two $\sigma$ from the synthetic spectrum. Outside the range of acceptable values (e.g. lower temperature values coupled with higher column densities or vice versa) no reasonable fit is found as the bandheads become less optically thick, and therefore they are more sensible to variations in $\textit {T}$ and $N_{\rm CO}$ (see Figs.\,\ref{fig:tau_T1800_NCO1e22}, and \ref{fig:tau_T2900_NCO9e20}).   
 
The observed spectrum is well reproduced by $v_{rot} \sin {i}$ values in the range of 130$\pm$30\,\kms. Different values of $v_{rot} \sin {i}$ change the width of the blue shoulder (especially for the first bandhead), as well as the location of the bandhead peaks (see Fig.\,\ref{fig:spec_err_vsini}). The range of acceptable values of $v_{rot} \sin {i}$ has been defined as that for which the synthetic and observed spectrum differ by less than two spectral channels. 

\begin{figure*}
    \centering
    \includegraphics[width=1.0\textwidth]{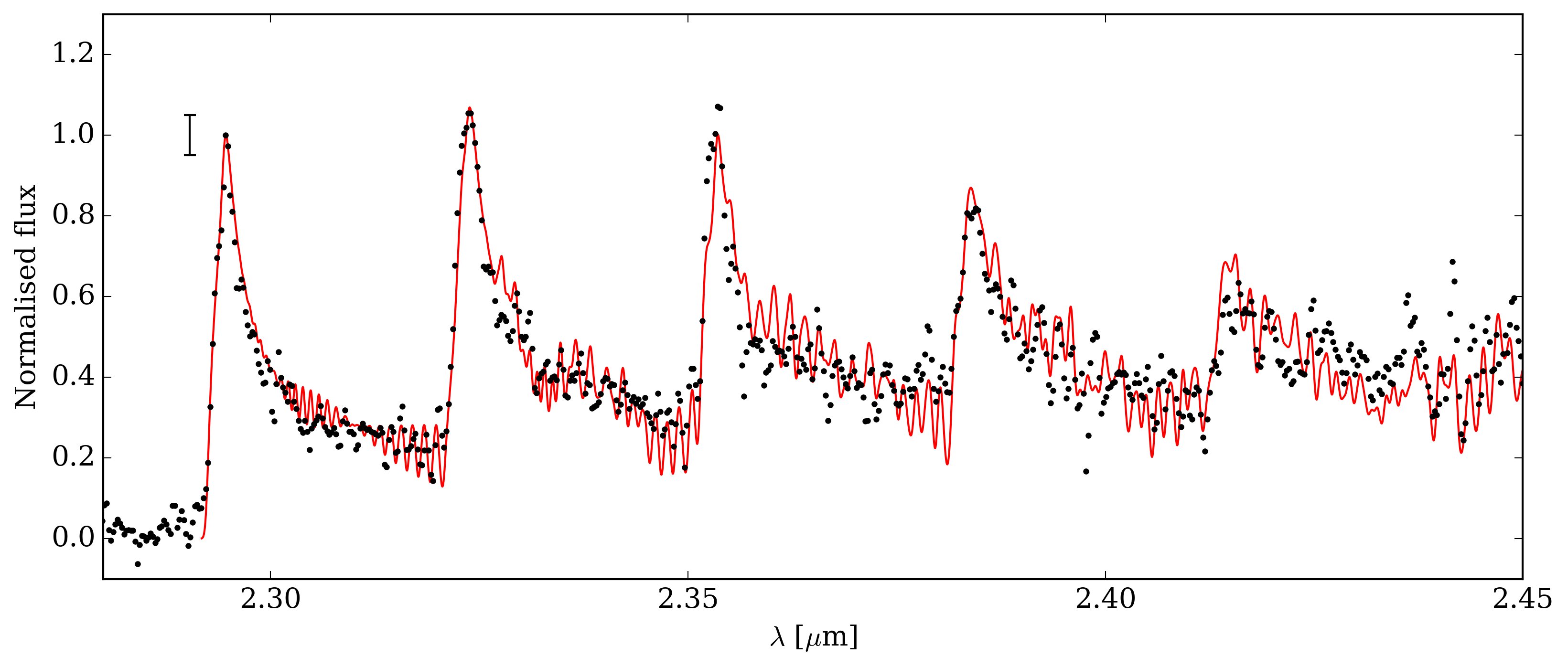}
    \caption{Continuum subtracted spectrum of 51 Oph around the CO bandhead emission for the May dataset (black dots) normalised to the peak of the first bandhead. Overplotted is the model spectrum (red solid line) computed for T =2400\,K, N$_{\mathrm{CO}}$=$4\times10^{21}\,\mathrm{cm}^{-2}$, \deltav=15\kms and \vsini = 130 \kms. The vertical black line at the upper left of the first bandhead represent twice the rms value measured in the observed continuum region adjacent to the first bandhead.}
    \label{fig:best_fit}
\end{figure*}

The intrinsic line width $\Delta \rm v$ of the single bandhead components is  a very interesting parameter related to the turbulence level. However, this parameter is very difficult to measure even at high spectral resolution, as its value is typically much lower than the line Keplerian broadening \citep{Lee2016}. In the case of 51\,Oph, we find acceptable values of $\Delta \rm v$ in the range of  $15 \pm 5$\,\kms\ (see Fig.\,\ref{fig:spec_err_dv}). 
As previously discussed, the individual $J$ components in the CO spectrum of 51\,Oph are very optically thick in the bandhead tails as well. This means that increasing $\Delta \rm v$ increases the emission in the line wings in such a way that, at the GRAVITY resolution, the bandhead tails look like a continuum of increasing value with respect to the peak. As done previously, the acceptable range of parameters was selected by modifying $\Delta \rm v$ until the observed spectrum (peak and/or tail of the bandheads) differed by more than two $\sigma$ from the synthetic spectrum. The best range of acceptable fitting parameters is shown in Fig.\,\ref{fig:best_fit} and reported in Table\,\ref{tab:best_fit_CO}.

\begin{table}[ht]
\caption{CO  properties derived from the spectrum fitting}  
\vspace{1.0 cm}
\label{tab:best_fit_CO}      
\centering                          
\begin{tabular}{c c c c}        
\hline\hline                 
Temperature &CO column density&\deltav&$v_{rot}\sin{i}$ \\ 
K& $\mathrm{cm^{-2}}$&\kms&\kms\\
\hline                        
    $1900-2800$&$(0.9-9)\times 10^{21}$&$15\pm 5$&130$\pm 30$\\      
   \hline                                   
\end{tabular} 
\end{table}

\subsection{Modelling the CO line displacement} 

\begin{table}[ht]
\caption{CO properties derived from the line displacement}  
\vspace{1.0 cm}
\label{tab:best_fit_CO_displacement}      
\centering                          
\begin{tabular}{c c c }        
\hline\hline                 
R      & $i$ & v$_{rot}$ \\ 
au & deg & \kms\\
\hline                        
0.10$\pm$0.02 & 70$\pm$10 & 140$\pm$30 \\      
   \hline                                   
\end{tabular} 
\end{table}

\begin{figure*}
    \centering
    \includegraphics[width=1.0\textwidth]{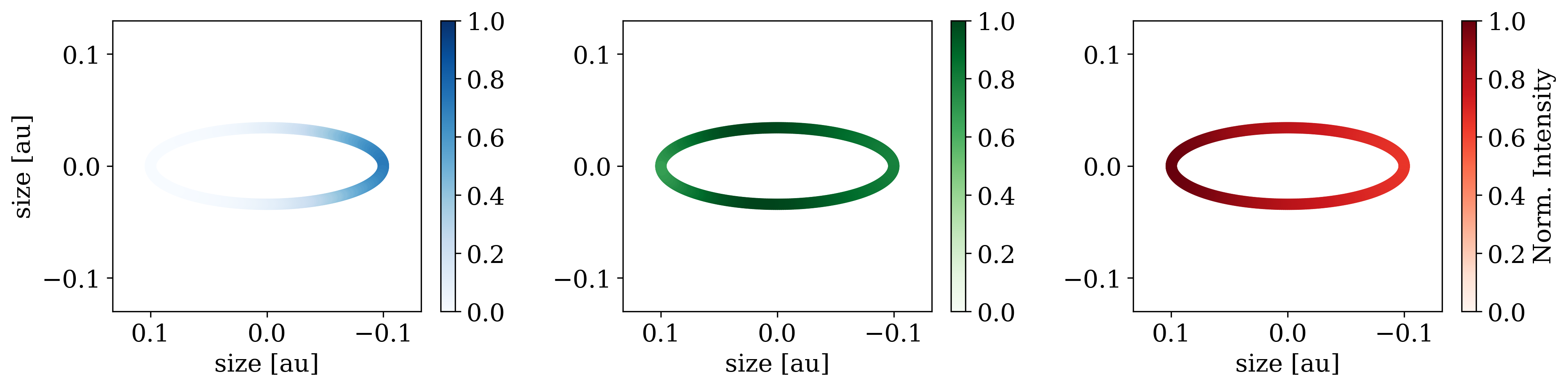}
    \includegraphics[width=1.0\textwidth]{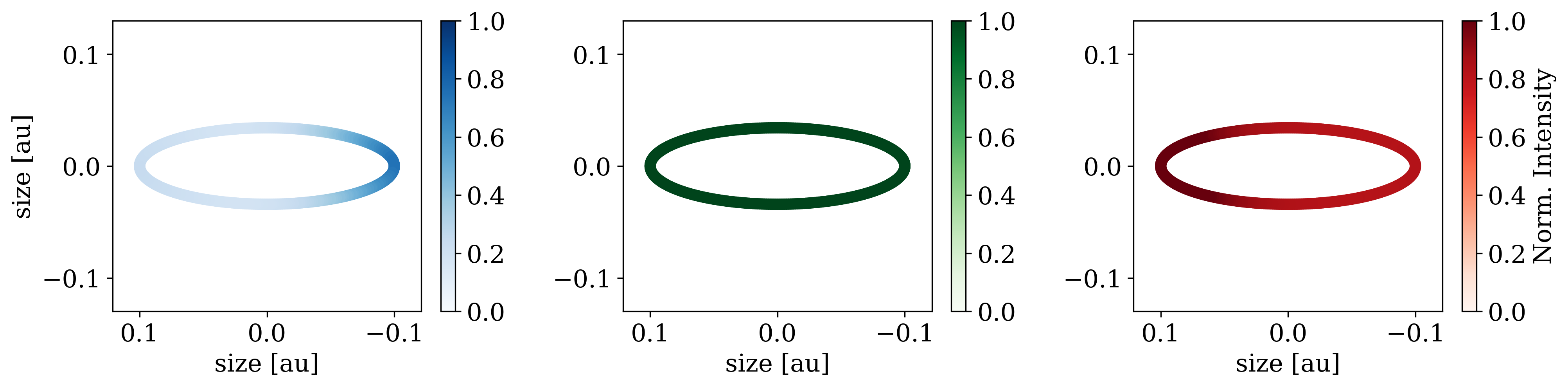}
    \includegraphics[width=1.0\textwidth]{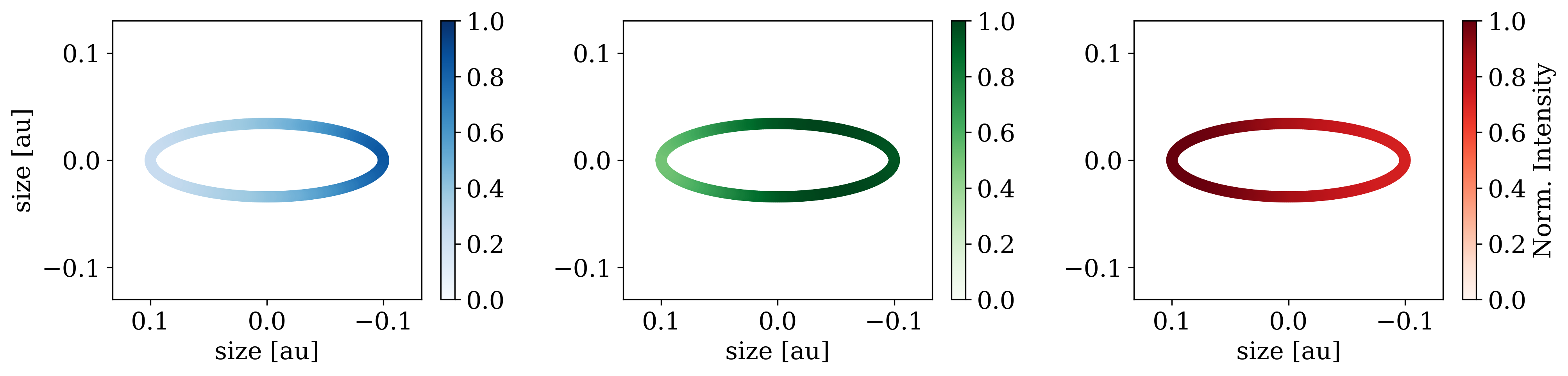}
    \caption{Model-predicted intensity maps of the blue-shifted (average over spectral channels at 2.2925\,\um, 2.3213\,\um, and 2.3516\,\um) , peak (average over spectral channels at 2.2945\,\um, 2.3237\,\um, and 2.3529\,\um), and red-shifted (average over spectral channels at 2.2954\,\um, 2.3246\,\um, and 2.3543\,\um) line components of the first (top panel), second (middle panel) and third (bottom) bandheads.
    Maps are normalised to the peak intensity and are displayed for  an inclination of 70\degr, and a radius of 0.1 au. The physical parameters of the model are the same as in Fig.\,\ref{fig:best_fit}.}
    \label{fig:int_maps}
\end{figure*}


To derive the synthetic values of the CO line displacement, $\delta$, the
azymuthal distribution of the CO intensity along the modelled ring  was computed as a function of wavelength for the first three bandheads. 
Examples of intensity maps computed for an inclination of 70\degr, a size of the CO emitting region of 0.1\,au, and the physical properties of T=2400\,K, \Nco=$4\times10^{21}$\,cm$^{-2}$, \deltav=15\,\kms, and \vsini=130\,\kms\ are shown in Fig.\,\ref{fig:int_maps} for the first three CO bandheads at a spectral resolution of $\mathcal{R}=4000$ and three velocities. 
As pointed out in Sec.\,3.4, only the blue-shifted emission is confined to a specific region of the ring (as would be the case for an individual line), while the emission is  much more homogeneous at all other wavelengths. It should be mentioned that as long as  $T$ and $N_{\rm CO}$ vary in a correlated way, the intensity distribution does not vary significantly within the range of acceptable values reported in Table\,\ref{tab:best_fit_CO}.

By inclining and rotating these intensity maps along the line of sight, synthetic $\delta$ values can be computed and used to further constrain the geometry and size of the CO emitting region. This was done by computing as a 
%
function of wavelength the photo-centre shifts with respect to the centre of the ring for the 12 observed PAs. 
In principle, the variation of $\delta$ with wavelength depends on the same physical parameters that reproduce the CO spectrum ($T$, $N_{\rm CO}$, and \deltav), and on the geometry (i.e. $i$ and PA), kinematics ($v_{rot}$) and size of the emitting region. We then started by creating an intensity map that reproduces the CO spectrum (see Table\,\ref{tab:best_fit_CO}), and exploring a range of sizes, inclinations, PA, and rotational velocities that produce a  good match between our observations and the model.
%

The resulting synthetic $\delta$ values estimated from the intensity maps shown in Fig.\,\ref{fig:int_maps} are plotted as solid black lines in Figs.\,\ref{fig:pshifts_model_obs_1st}, \ref{fig:pshifts_model_obs_2nd}, and \ref{fig:pshifts_model_obs_3rd} along with the $\delta$ values derived from the observations (see Sect.\,3.4) for the first three bandheads.
To interpret our results, we need to take into account that for an inclined ring in Keplerian rotation, the greatest shift in velocity is expected to occur along the major axis of the projected ring, while the smallest shift (close to zero) is expected along the minor axis.
This is exactly what is seen for the bluest observable wavelengths of each single bandhead as this wavelength range behaves similarly to an individual line (see Figs.\,\ref{fig:pshifts_model_obs_1st} to \ref{fig:pshifts_model_obs_3rd}). However, the mixture of blue- and red-shifted individual $J$ components at the peak and the tails of the bandheads results in very small photo-centre asymmetries with respect to the continuum. The observed bluest $\delta$ values show maximum values along PA of 109\degr\ and 135\degr, indicating that the PA of the major axis of the CO emitting region must be oriented within this PA range. This is consistent with the PA=116\degr\ of the major axis of the dusty disk as derived from the continuum interferometric data. We therefore fixed the CO ring PA to that of the disk continuum emission, that is, we assumed that the CO and disk continuum emission have a similar PA within our uncertainties.

Following the same principle, the size of the CO emitting region can be accurately derived from the maximum displacement measured along the major axis of the disk.  $\delta$ always reaches its maximum value along the major axis. This value is independent of $v_{rot}$ and $i$ variations (see Figs.\,\ref{fig:shifts_h_res_diff_Vrot} and \ref{fig:shifts_PA180_h_res_inc}). By varying the size of the CO ring and comparing the modelled and observed displacement, we derive a size of the CO emitting region of 0.10$\pm$0.02\,au (see Fig.\,\ref{fig:shifts_diff_Radius}).

On the other hand, the displacement along orientations that are not aligned with the major axis PA can give important constraints on $v_{rot}$ and $i$. As shown in Fig.\,\ref{fig:shifts_P180_R4000_diff_Vrot}, lower rotation velocities produce smaller displacements and change the shape of the displacement as a function of wavelength. Similarly, lower inclinations produce maximum displacements,  which also modifies the shape of the displacement curve (Fig.\,\ref{fig:shifts_P135_R4000_diff_inc}). Following this, we found that the observed displacements along orientations that are not aligned with the major axis PA are well reproduced assuming a $v_{rot}= 140 \pm 30$ km/s and $i =70^\circ \pm 10$\degr. 

\section{Discussion}

Our study of the CO overtone bandheads in the 2.3\,\um\ region using GRAVITY shows that the CO properties are consistent with the  emission of a ring  of warm CO, located at a distance of $\sim$0.1\,au from the star, well inside the dust sublimation radius. 
The emitting region is likely quite narrow. The nominal width, derived by comparing the observed flux to the model emissivity at the peak of the first bandhead is (${\Delta R}/{R}=0.15$). This may be a lower limit, and some contribution from a cooler, more extended region with slightly lower rotational velocity cannot be ruled out. However, this very simple model is able to reproduce the basic features of our interferometric observations within an acceptable range of physical parameters. A more complex disc model with a larger set of free parameters is therefore not required to reproduce our observations, especially at our moderate spectral resolution.


51\,Oph is one of the few stars that exhibit bright ro-vibrational CO overtone emission, and it has been studied extensively in the past. Two spectroscopic studies using high spectral resolution data  have been performed in the past (\citealt{Thi2005}, at R=10000, and \citealt{Berthoud2007}, at R=25000). 
The observed spectra show only small differences, mostly due to the different spectral resolution. Both teams modelled the CO emission as coming from a  disc region with a radial temperature and column density gradient, and fit only parts of the emission spectrum, focusing on the first bandhead.  Both studies conclude that the CO is mostly emitted in a narrow, dust-free portion of the disc, very close to the central star.
In particular, \citet{Thi2005} fit the spectrum with a CO disc extending between 0.15 and 0.35 au. The temperature and CO column densities are T=$2850 \pm 500$\,K and $N_{\mathrm{CO}}$=(0.17-2.5)$\times 10^{20}$ cm$^{-2}$ at the inner radius, and  $T=1500$\,K and $N_{\mathrm{CO}}$\ 2.3 times lower at the outer radius, respectively. The lower temperature and CO column density at the outer radius make a strong contribution of the outer region to the CO total emission very unlikely, even when the emitting area is increased. This means that the outer CO disc region might mostly affect the band tail emission. On the other hand, \citet{Berthoud2007} allowed for a much larger number of free parameters (10). Their results show an emitting region with a ratio of $\sim 4$ between the outer and inner radii. The maximum temperature is very high ($\sim 4000$ K) and decreases steeply to $\sim 700$ K at the outer disc radius. The column density at the inner radius is $\sim 3 \times 10^{20} \mathrm{cm}^{-2}$. In this model, different portions of the spectrum require rather different parameter values. However, about 80\% of the flux  comes from the inner half of the disc, with an average temperature of 2700\,K and column density of $7.5 \times 10^{20} cm^{-2}$. In both studies, the values reported at the inner CO disc radius, where most of the CO emission is most likely generated, are similar to those found in our ring modelling 
and confirm our main conclusions: most of the CO emission comes from a relatively warm and dense narrow region (see Table\,\ref{tab:best_fit_CO}), very close to the star, well inside the dust sublimation radius.

 
Regarding the geometrical constraints, the disc inclination is poorly constrained by spectroscopic fittings alone, with a preference for very high values ($\sim 88 $\degr).
In a pioneering paper, \citet{Tatulli2008} used the visibilities from a single observation with the ESO VLTI instrument AMBER using the 8m UT telescopes to derive a size of the CO emission of 0.15\,au and a disc inclination of  85$\pm ^{+5}_{-15}$\degr.
Our interferometric observations and the differential phase signal in particular give a more moderate value of $i= 70^\circ \pm 10$ \degr , but this is still consistent within the uncertainties with all previous estimates.

The  overtone CO emission provides one of the very few direct tests of the physical and chemical properties of the innermost gaseous discs. It is unfortunate that no detailed model exists for this region, and we can only be guided by considerations based on simple accretion disc models. \citet{Muzerolle2004} computed models for the inner disc around a star of $T\sim 9000$\,K and $L=30$\,\Lsun, following the procedure of solving the radiation transfer as outlined in \citet{Calvet1991}. They used mean opacities, computed assuming local thermal equilibrium, which are very likely not appropriate for the disc atmosphere, but provide a solution for the disc midplane temperature \citep{Dullemond2010}. 
At 0.10\,au from the star, the midplane disc temperature for 51\,Oph is around 2000\,K assuming T$_{\mathrm{eff}}$=10000\,K,  somewhat lower than the CO temperature.  The column density of gas at the distance $R$ from the star can be written as a function of the midplane temperature, the mass accretion rate, and the viscosity parameter $\alpha_\mathrm{{visc}}$ \citep{Lynden-Bell1974}. The mass accretion rate of 51 Oph is not well known and is probably lower than  $< 10^{-7}$\,\msyr\ \citep[][and discussion therein]{Mendigutia2011}.  For a nominal value $\alpha_{visc}=0.01$, the expected gas column density is therefore $N_H < 4.6\times 10^{27} \rm{cm}^{-2}$.  Assuming a CO/H of $\sim 10^{-4}$, the total column density of warm gas required by the observed CO emission is $\sim 4 \times 10^{25} \rm{cm}^{-2}$, and the CO emitting region is located above the midplane, at $z \sim 2.6 H_p$, where $H_p$ is the pressure scale height. The density in this region is very high, about $3 \times 10^{15} \rm{cm}^{-3}$, sufficient to thermalize the CO vibrational levels \citep{Thi2013}.  For $M_{acc}=10^{-8}$\,\msyr, the CO emitting region comes closer to the midplane, at $z\sim 1.7 H_p$, with similar density. It is therefore likely that the CO is emitted in a column of gas slightly above and somewhat warmer than the midplane. Similar results were observed in the case of MYSO \citep[see][for more details]{CarattioGaratti2020}.

This column of warm CO in the intense radiation field that characterises the inner disc of 51 Oph depends on a number of chemical and physical processes, that need to be understood in detail. Recently, \cite{Bosman2019} suggested that CO is unlikely to exist in dust-depleted  regions of Herbig Ae stars. However, the models do not extend to the innermost dust-free disc regions and have been computed for lower luminosity stars ($\sim$30\,\Lsun; the luminosity of 51\,Oph is 250 \Lsun). The detections and characteristics of the vibrationally excited CO in 51\,Oph (and of a few other HAeBe stars) indicate the need of computing appropriate models for the innermost dust-free regions and explore their robustness. Work in this direction is currently in progress (Gorti et al. in prep.). 



On the other hand, as mentioned in the introduction, a detection of the overtone CO emission in Class II stars is rare.  To the best of our knowledge, in addition to 51\,Oph, it has been detected in 9 of the $\sim$ 91 objects surveyed by \citet{Ilee2014}. Although the small number of detections prevents any statistical analysis, the lack of detections at luminosities below $\sim$20\,\Lsun\ and \macc\ below $\sim$ 10$^{-8}$\,\Msun/yr is conspicuous. This is confirmed by the very low number of detections in T Tauri stars, where only three of the many objects searched so far show CO emission (including the whole Lupus sample studied by \citealt{Alcala2017}; see \citealt{Koutoulaki2019}). Only one detection of the 15 objects has \Lstar $> 16000$ \Lsun. Within this wide parameter range, the CO detections are distributed  rather uniformly. 
 These results indicate that a low stellar or accretion luminosity and/or a low mass-accretion rate prevent the formation of a sufficiently high column density of warm CO. This is confirmed by the detection of the overtone CO emission in some EXors during  outbursts only \citep[e.g. ][]{Aspin2010,Zhen2020}. 
  High-mass young objects tend to have a larger fraction of CO detections than lower luminosity ones. Most recently, \cite{Pomohaci2017} detected the CO overtone emission in a subset of 7 out of 38 massive young stellar objects (MYSO). These detections seem to be confined to a narrow range of \macc\ ($10^{-4.5}$ -- $10^{-3.0}$\,\msyr), where they account for about 40\% of the objects. 
  As for lower luminosity objects, most upper limits to the CO bandhead luminosity are well below the detections, suggesting that the relatively small fraction of detections is not a sensitivity issue.
  
  However, the detection of CO emission in only a few objects among many of similar properties remains puzzling. 
  It is possible that overtone CO emission occurs only in a very limited range of physical conditions in the inner dust-free regions of discs because of a combination of factors, such as mass accretion rates, stellar and accretion luminosity, and X-ray emission from the central star that can only be understood through detailed non-local thermal equilibrium models. Even if rare, the detection and characterisation of this emission can give us the only constraints on the innermost gaseous regions of discs, in which accretion onto the star occurs and from which winds are launched.

\section{Conclusion}
We reported our results on the Herbig Ae/Be star, 51\,Oph, using VLTI/\textit{GRAVITY} interferometric observations in the K band at high spectral resolution. Our main conclusions are listed below.

\begin{itemize}
    \item 51 Oph shows prominent CO bandhead emission. The first four ro-vibrational transitions are clearly detected in the spectrum, as well as an indication of the fifth. Our GRAVITY observations show a clear increase in visibility signal within the first four bandheads, indicating that the CO emitting region is more compact than the total continuum emission. Clear differential phase signatures are observed at the positions of the first four bandheads.
    \item By fitting the continuum visibilities, we derive a stellar radius of $R_*$=0.4$\pm$0.1\,mas ($R_*$=10.6$\pm$2.6\,\Rsun) and a dusty disc with radius $R_{disc}$=4.0$\pm$0.8\,mas ($R_{disc}$=0.50$\pm$0.01\,au) at an inclination and PA of 63\degr$\pm$1\degr\ and 116\degr$\pm$1\degr. 
    \item By modelling the CO bandhead emission, we created intensity maps of the CO emission that were used to reproduce the observed spectrum and the differential phase signatures to broadly constrain the physical properties (T and $N_{\mathrm{CO}}$), kinematics ($v_{rot}$) and geometry (size, $i$, PA) of the CO emitting region. 
    \item Our modelling shows that the bandhead CO emission in 51 Oph is due to an overlap of optically thick individual J components. This makes it difficult to accurately constrain the T and $N_{\mathrm{CO}}$, as a change in T can be compensated for by a change in $N_{\mathrm{CO}}$. Nevertheless, in order to reproduce the CO spectrum a warm (T$\sim$1900--2800\,K) and dense  (10$^{20}$--10$^{21}$\,cm$^{-2}$) gas is needed in agreement with previous results \citep[e.g. ][]{Thi2013_51Oph, Berthoud2007}.
    \item The synthetic intensity maps of the CO emission were used to reproduce the observed CO displacements with respect to the photocentre of the continuum. From the modelling and analysis of the observed displacement an estimate of the size, inclination, PA, and rotational velocity of the CO emitting region was derived. A CO ring of radius $R_{CO}\sim$0.1\,au reproduces the maximum displacement at the bluest wavelength of the three first bandheads well. In addition, we find that the the CO PA and inclination is consistent with that of the dusty disk, indicating no major misalignment between the dusty disc and the CO emitting region. 
\end{itemize}

\begin{acknowledgements}
 We would like to thank the referee for his/her fruitful comments and suggestions. It significantly improved the manuscript. M.K. is funded by the Irish Research Council (IRC), grant GOIPG/2016/769, SFI grant 13/ERC/12907, and DFG grant FOR2634/1TE1024/1-1. R.G.L has received funding from Science  Foundation  Ireland under Grant No. 18/SIRG/5597. R.F. acknowledges support from Science Foundation Ireland (grant No. 13/ERC/12907). A.C.G. and T.P.R. have received funding from the European Research Council (ERC) under the European Union's Horizon 2020 research and innovation programme (grant agreement No.\ 743029). A.N. acknowledges the kind hospitality of DIAS. A.A., M.F. and P.G. were supported by Funda\c{c}\~{a}o para a Ci\^{e}ncia e a Tecnologia, with grants reference UIDB/00099/2020 and SFRH/BSAB/142940/2018.

\end{acknowledgements}

\bibliographystyle{aa}
\bibliography{references.bib}{}

 \begin{appendix}

\section{CO spectrum}

In this section we illustrate with examples how the spectrum of the CO bandhead emission changes when different free parameters are varied. We start by showing that the CO bandhead emission in 51\,Oph is due to an overlap of optically thick individual J components (Sect.\,\ref{sect:tau}), followed by a description of the spectral variation when T and \Nco\ are varied (Sect.\,\ref{sect:T_Nco}), and the intrinsic line width and \vsini\ values (Sect.\,\ref{sect:line_width}). 

\subsection{Optical depth}
\label{sect:tau}

\begin{figure*}
     \centering
     \includegraphics[width=\linewidth]{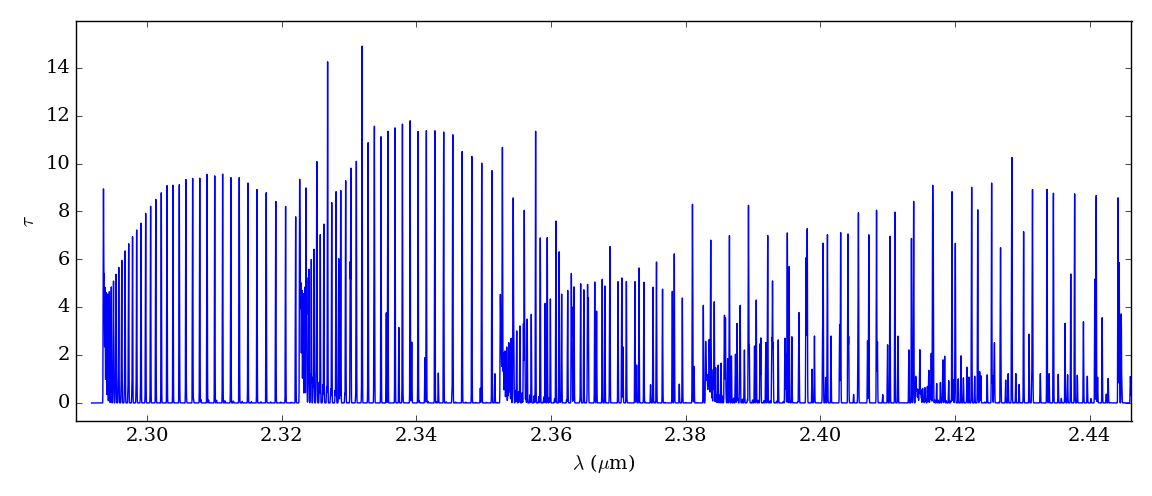}
     \caption{Optical depth vs. wavelength for the first five CO bandheads ($\upsilon$=2--0, 3--1, 4--2, 5--3, and 6--4, ) at R=100\,000. The CO is modelled as a ring at T=2400\,K, $N_{\mathrm{CO}}$=4$\times$10$^{21}$\,cm$^{-2}$, and \vsini=130\,\kms. The intrinsic CO line width is \deltav=15\,\kms.}
     \label{fig:tau_plot}
 \end{figure*}

\begin{figure*}
    \centering
    \includegraphics[width=1.0\textwidth]{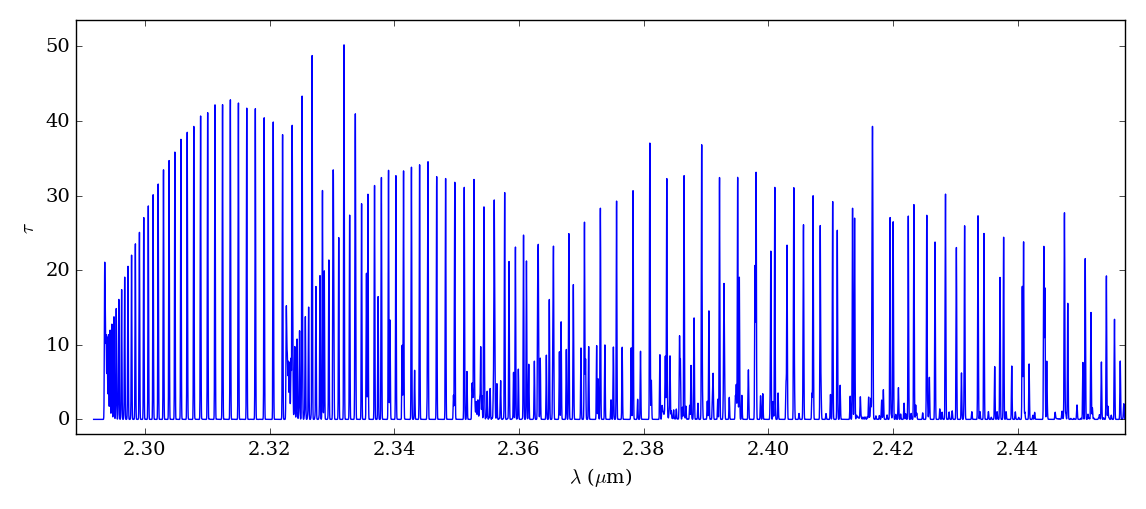}
    \caption{
    Same as Fig.\,\ref{fig:tau_plot}, but for T=1800\,K and $N_{\mathrm{CO}}$=10$^{22}$\,cm$^{-2}$.}
    \label{fig:tau_T1800_NCO1e22}
\end{figure*}

\begin{figure*}
    \centering
    \includegraphics[width=1.0\textwidth]{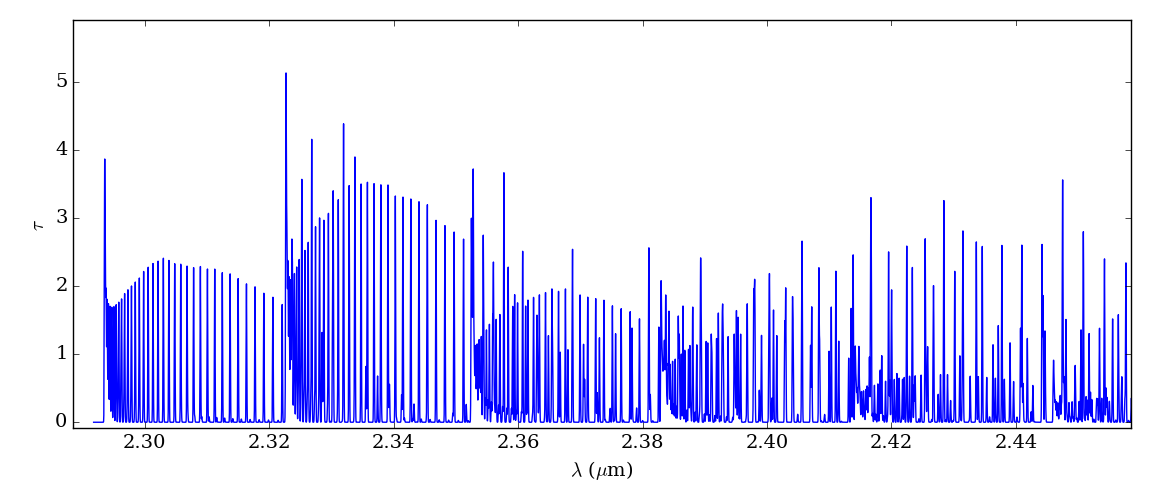}
    \caption{
    Same as Fig.\,\ref{fig:tau_plot}, but for T=2900\,K and $N_{\mathrm{CO}}$=9$\times$10$^{20}$\,cm$^{-2}$.}
    \label{fig:tau_T2900_NCO9e20}
\end{figure*}

The CO bandhead emission in 51 Oph is due to an overlap of optically thick individual J components. The two last bandheads ($\upsilon$=5--3 and 6--4) are optically thinner than the first three bandheads ($\upsilon$=2--0, 3--1, and 4--2). This is illustrated in Fig.\,\ref{fig:tau_plot}, where the optical depth ($\tau$) of the first five CO bandheads versus wavelength is plotted at a spectral resolution of R=100\,000. As a consequence, the temperature and CO column densities cannot be accurately constrained because an increase or decrease in T can be compensated for by an opposite change in column density. Figures\,\ref{fig:tau_T1800_NCO1e22} and \ref{fig:tau_T2900_NCO9e20} illustrate that an increase (decrease) in temperature can be compensated for by a decrease (increase) in CO column density so that most of the individual J components remain optically thick. When T is decreased and $N_{\mathrm{CO}}$ is increased the optical depth increases from the most strongly red-shifted towards the most strongly blue-shifted wavelengths (Fig.\,\ref{fig:tau_T1800_NCO1e22}). Eventually, all the J-components reach the equivalent Planck function. In contrast, a gradual increase and decrease of T and $N_{\mathrm{CO}}$, respectively, decreases the optical depth from the highest to the lowest $\upsilon$ band components until they eventually become optically thin (Fig.\,\ref{fig:tau_T2900_NCO9e20}).

%
%

\subsection{Temperature and column density}
\label{sect:T_Nco}

\begin{figure*}
     \centering
     \includegraphics[width=\linewidth]{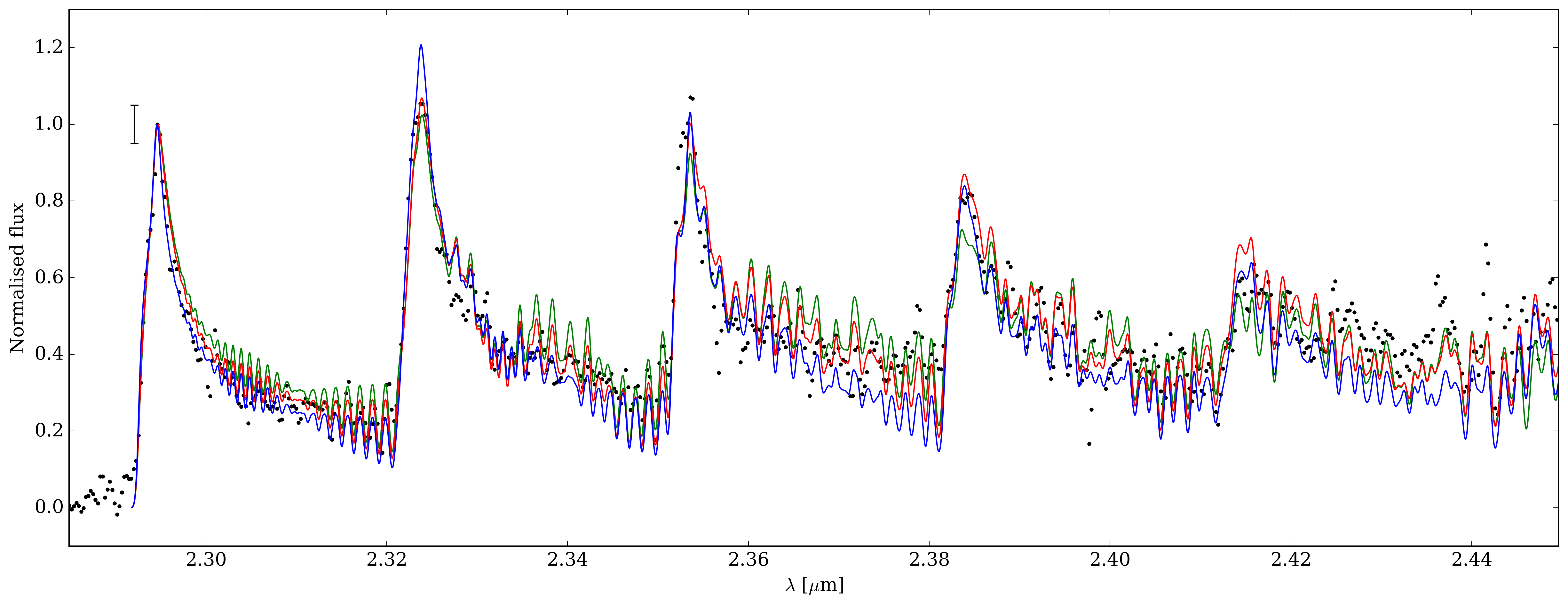}
     \caption{Continuum-subtracted spectrum of 51\,Oph normalised to the peak intensity of the first bandhead (black dots). Predictions of theoretical CO ring models as described in the main text are overplotted (solid coloured lines). The vertical black line in the upper left corner of the first bandhead represent twice the rms value measured in the observed continuum region adjacent to the first bandhead. The models have been computed for combinations of T and \Nco\ values of T=1800\,K, \Nco=$1\times 10^{22}\, \mathrm{cm}^{-2}$ (green), T=2400\,K, \Nco=$4\times 10^{21}\,\mathrm{cm}^{-2}$ (red), and T=2900\,K, \Nco=$8\times 10^{20}\,\mathrm{cm}^{-2}$ (blue) with a constant value of \deltav=15\,\kms, and \vsini=130\,\kms.}   
     %
     \label{fig:spec_fit_ddif_T_NCO}
 \end{figure*}

\begin{figure*}
     \centering
     \includegraphics[width=\linewidth]{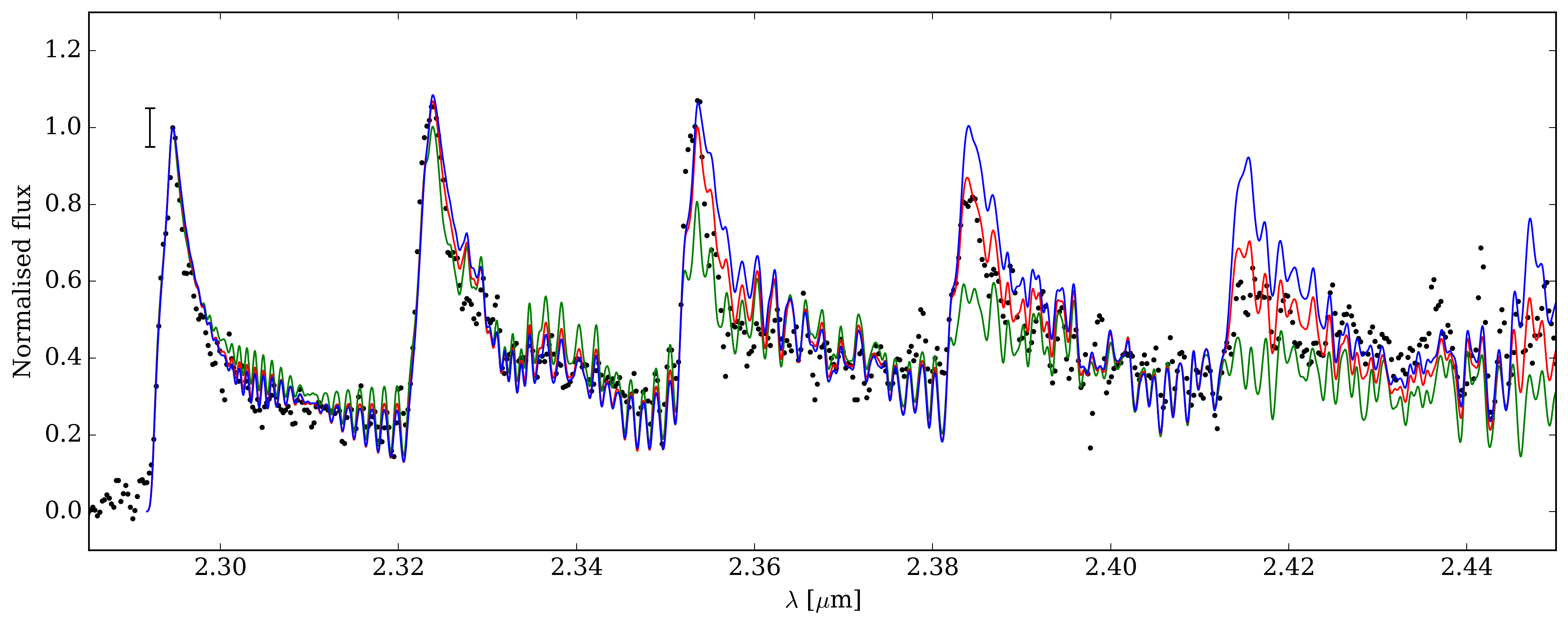}
     \caption{Same as Fig.\,\ref{fig:spec_fit_ddif_T_NCO}, but for models with fixed values of \deltav= 15\,\kms, \vsini= 130\,\kms, and \Nco= $4\times10^{21}$\,cm$^{-2}$, and different  temperature values of T=1800\,K (solid green line), T=2400\,K (solid red line), and T=2900\,K(solid blue line).}
     \label{fig:spec_err_T}
 \end{figure*}

\begin{figure*}
     \centering
     \includegraphics[width=\linewidth]{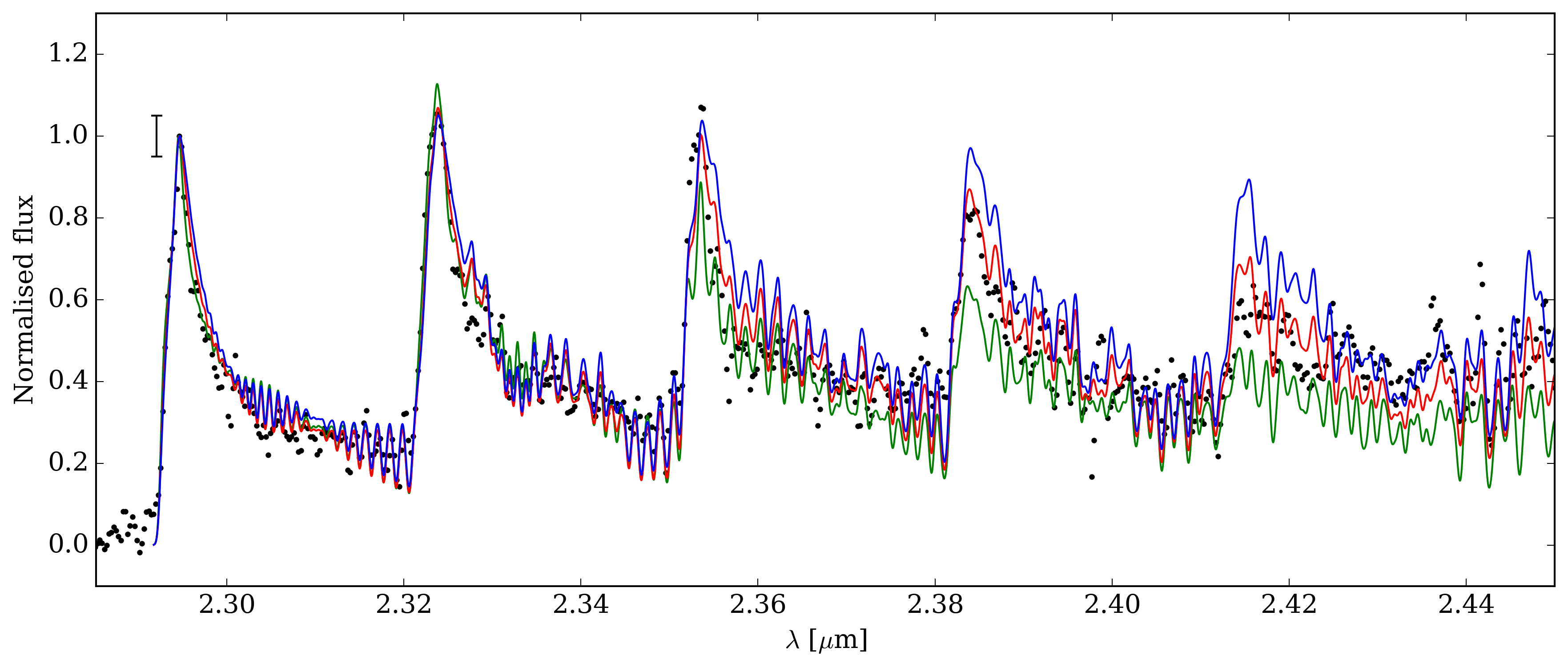}
     \caption{Same as Fig.\,\ref{fig:spec_err_T}, but for fixed values of \deltav= 15\,\kms, \vsini= 130\,\kms, and T= 2400\,K, and different column density values of \Nco= $8\times 10^{20}\,\mathrm{cm}^{-2}$ (solid green line), \Nco= $4\times10^{21}\,\mathrm{cm}^{-2}$ (solid red line), and \Nco= $1\times 10^{22}\,\mathrm{cm}^{-2}$ (solid blue line).}
     \label{fig:spec_err_NCO}
 \end{figure*}

Figure\,\ref{fig:spec_fit_ddif_T_NCO} shows the observed continuum subtracted spectrum of 51\,Oph normalised to the peak of the first bandhead, along with the modelled CO spectrum for the same parameters discussed in Sect.\,\ref{sect:tau}: CO is emitted by a ring in Keplerian rotation with \vsini=130\,\kms, and combinations of temperature and CO column densities of 1800\,K, 1$\times$10$^{22}$\,cm$^{-2}$; 2400\,K, 4$\times$10${21}$\,cm$^{-2}$; and 2900\,K, 9$\times$10$^{20}$\,cm$^{-2}$. Both the synthetic and observed spectra are shown at R=4000. Our observations are relatively well reproduced by a wide range of T and \Nco\ values because the bandheads are due to an overlap of mostly optically thick J components (Figs.\,\ref{fig:tau_plot} to \ref{fig:tau_T2900_NCO9e20}).
For comparison, Figs.\,\ref{fig:spec_err_T} and \ref{fig:spec_err_NCO} show the change caused by single variations of T and \Nco\ in the CO bandhead line profiles when the other free parameters are fixed.

 
 \subsection{Intrinsic line width and \vsini}
 \label{sect:line_width}
 
\begin{figure*}
     \centering
     \includegraphics[width=\linewidth]{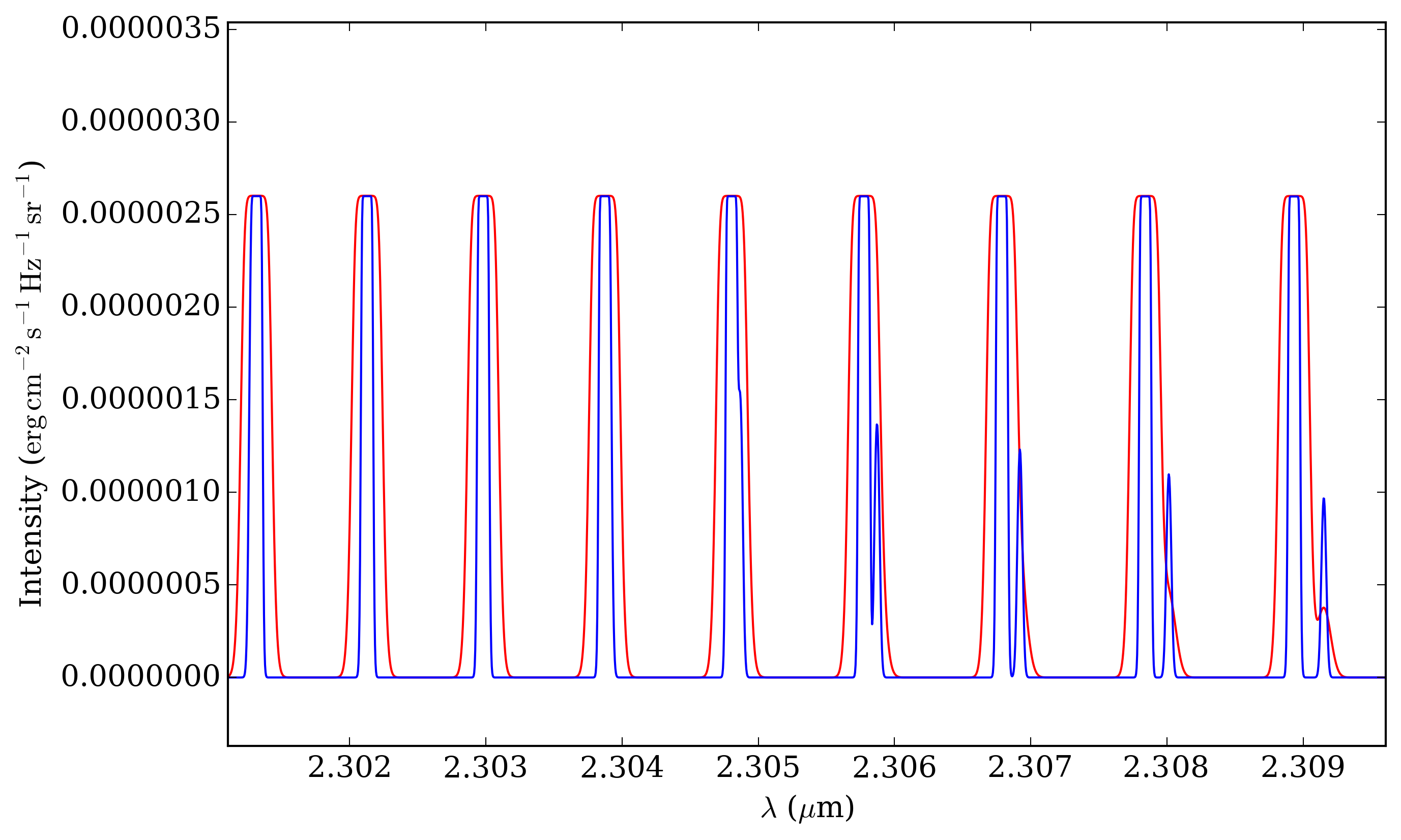}
     \caption{Individual CO J components profiles contributing to the line intensity of the first CO bandhead ($\upsilon$=2--0) for wavelengths ranging from 2.302\,\um\ to 2.309\,\um\ at a spectral resolution R=100\,000. Models computed for intrinsic line widths of \deltav=5\,\kms\ (blue solid line), and \deltav=15\,\kms\ (red solid line), and fixed values of T=2400\,K, \Nco=$4\times10^{21}$\,cm$^{-2}$, and \vsini=130\,\kms\ are shown. To facilitate comparison, the model with \deltav=5\,\kms\ has been scaled-up to match the peak intensity of the J components with higher \deltav\ value.}
     \label{fig:j_comp_dv}
 \end{figure*}

\begin{figure*}
     \centering
     \includegraphics[width=\linewidth]{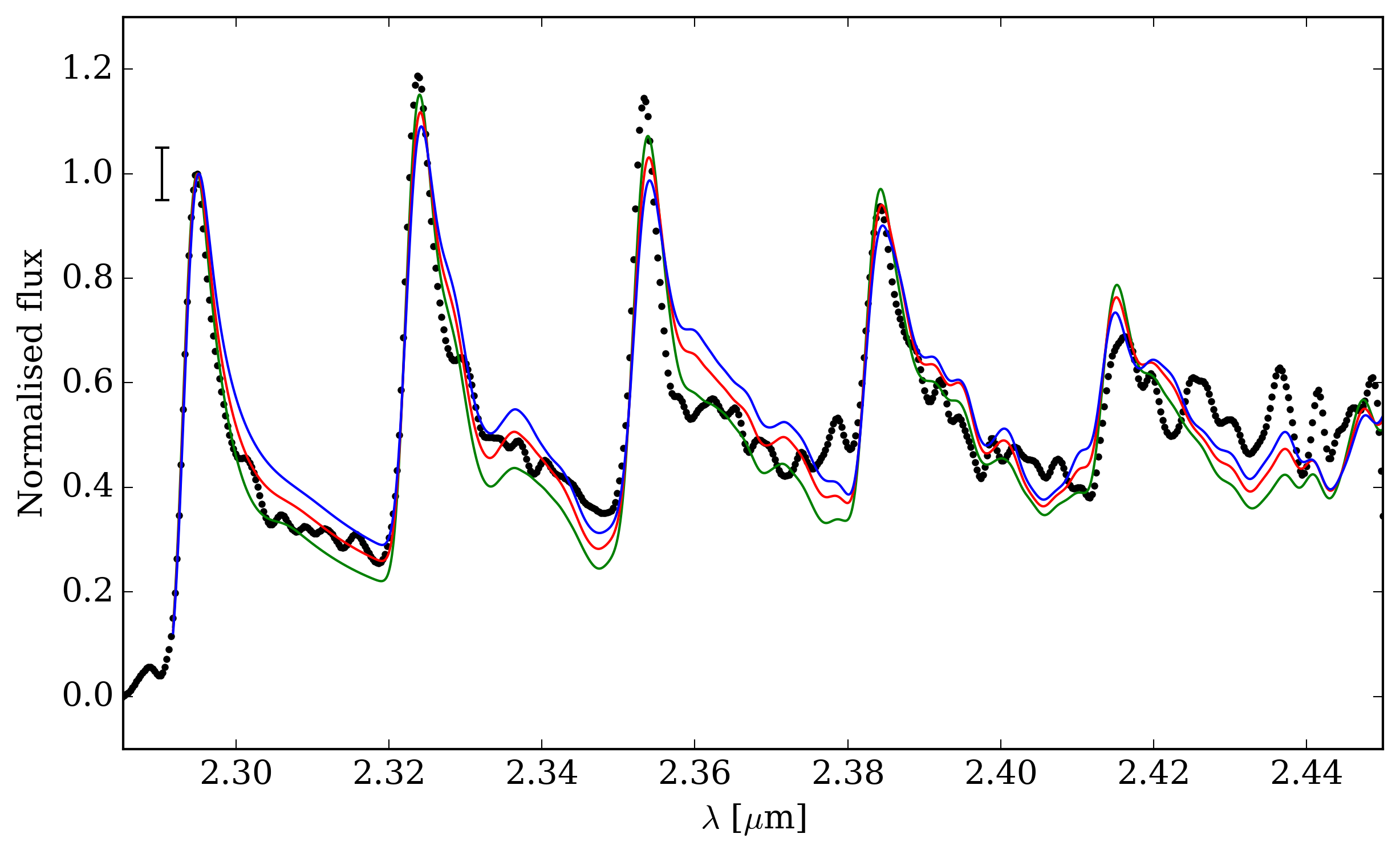}
     \caption{Same as Fig.\,\ref{fig:spec_fit_ddif_T_NCO}, but for models with fixed values of T=2400\,K, \Nco=$4\times10^{21}$\,cm$^{-2}$, and \vsini=130\,\kms. Coloured solid lines represent models with different values of the intrinsic line width: \deltav= 10\,\kms\ (solid green line), \deltav= 15\,\kms\ (solid red line), and \deltav= 20\,\kms\ (solid blue line).The data are smoothed at a resolution of R=1000. When the data are smoothed the first bandhead becomes smaller because the line peak has only a few points, which causes the second and third bandhead to exceed than the unsmoothed spectra when we normalise. }
     \label{fig:spec_err_dv}
 \end{figure*}

 In addition to the physical parameters, the CO bandhead profiles also depend on the assumed intrinsic line width of the individual J components. 
 To better illustrate this dependence, we produced very high spectral resolution spectra (R=100\,000) of the individual J components contributing to the tail of the first bandhead in the wavelength range between 2.302\,\um\ to 2.309\,\um\ for two different \deltav\ values (see Fig.\,\ref{fig:j_comp_dv}). This figure shows that an increased \deltav\ value broadens the profile of the J components while the emitting area remains the same. As a consequence, an increase in \deltav\ produces a blend of J components. 
 This effect translates into variations in line flux of the CO bandheads as shown in Fig.\,\ref{fig:spec_err_dv}. This figure shows the continuum-subtracted spectrum of 51\,Oph (black dots) normalised to the peak of the first bandhead, along with three different synthetic spectra. The modelled spectra were computed for the same T, \Nco, and \vsini, but the intrinsic line width was varied from 10\,\kms\ to 20\,\kms. The increased blend of J components for higher values of \deltav\ results in a gradual increase in the bandhead flux.
 
 %
 %
 
%
%
 
\begin{figure*}
     \centering
     \includegraphics[width=\linewidth]{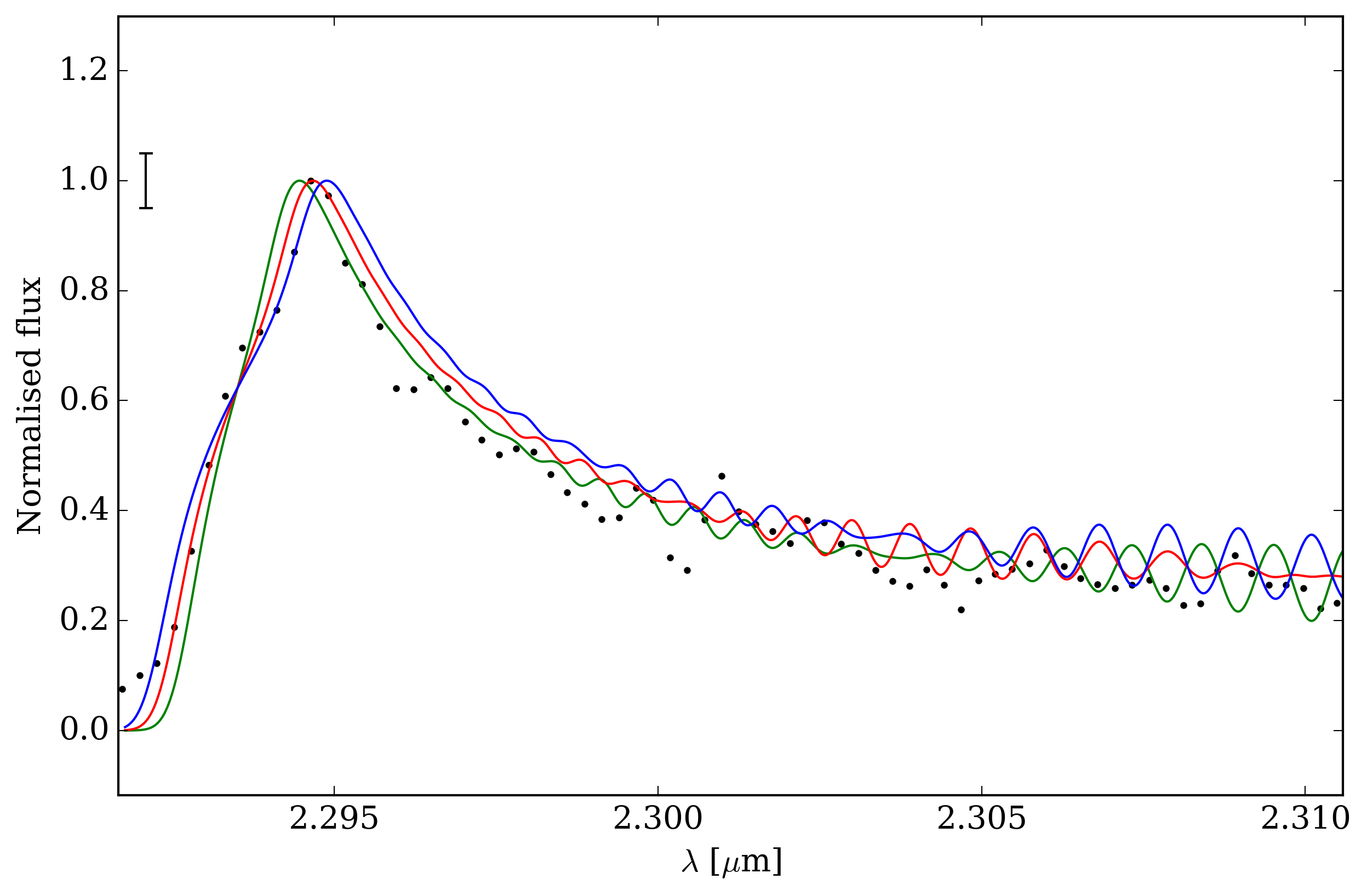}
     \caption{Continuum-subtracted first CO bandhead of 51\,Oph normalised to the peak intensity (black dots). Coloured lines shown models with fixed values of T=2400\,K, \Nco=$4\times10^{21}$\,cm$^{-2}$, \deltav=15\,\kms and \vsini\ values of 100\,\kms\ (solid green line), 130\,\kms\ (solid red line), and 160\,\kms\ (solid blue line). The black vertical line in the top left corner of the peak of the bandhead has a lengths of twice the rms value measured at the continuum.}
     \label{fig:spec_err_vsini}
 \end{figure*}

 On the other hand, changes in \vsini\ also produce changes in the CO banhdead profiles. At our moderate spectral resolution, changes in \vsini\ mostly affect the blue-shifted shoulder of the CO first bandhead and its peak position. Higher values of \vsini\ produce more pronounced blue-shifted shoulders, and shift the peak of the bandhead towards red-shifted wavelengths (see Fig.\,\ref{fig:spec_err_vsini}).
 
\section{CO line displacements}

 The CO line displacement with respect to the continuum photo-centre depends on the physical properties of the CO (T, \Nco, and \deltav), its geometry ($i$, PA), kinematics ($v_{rot}$), and on the size of the emitting region. For the particular case of 51\,Oph and the moderate spectral resolution of our observations, the line displacement does not significantly change within the range of acceptable values of T, \Nco, and \deltav, but it can help to better constrain the $i$, PA, and size of the CO emitting region.

\begin{figure*}
    \centering
    \includegraphics[width=1.0\textwidth]{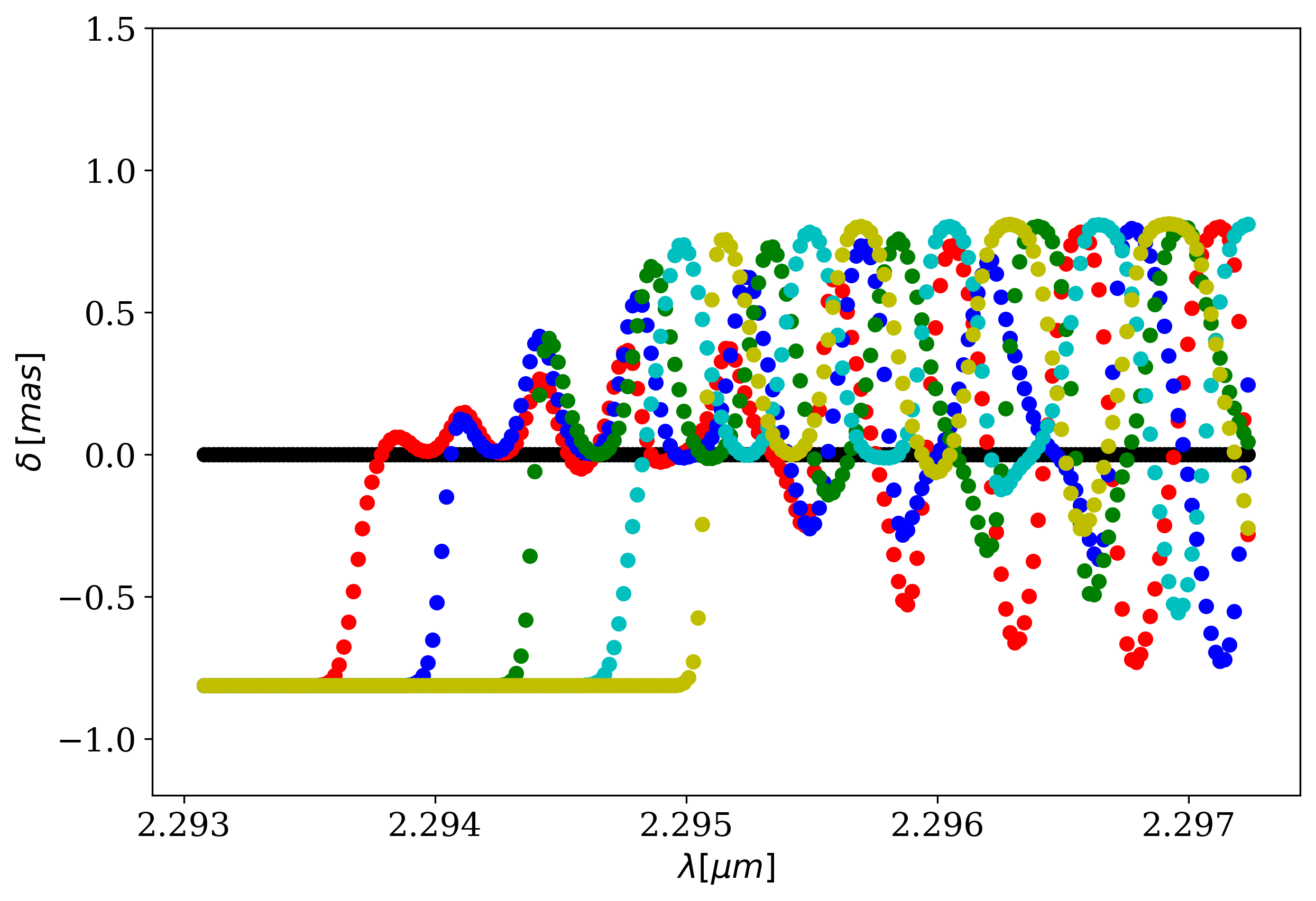}
    \caption{Synthetic displacements across the first CO bandhead at R=50\,000 along the major axis (PA=180\degr) of a ring in Keplerian rotation 0.2\,au across, at $i$=70\degr, and $v_{rot}$ ranging from 0\,\kms\ (black) to 250\,\kms\ (green-yellow) in steps of 50\,\kms\ (red, blue, green, and cyan).   
    }
    \label{fig:shifts_h_res_diff_Vrot}
\end{figure*}

\begin{figure*}
    \centering
    \includegraphics[width=1.0\textwidth]{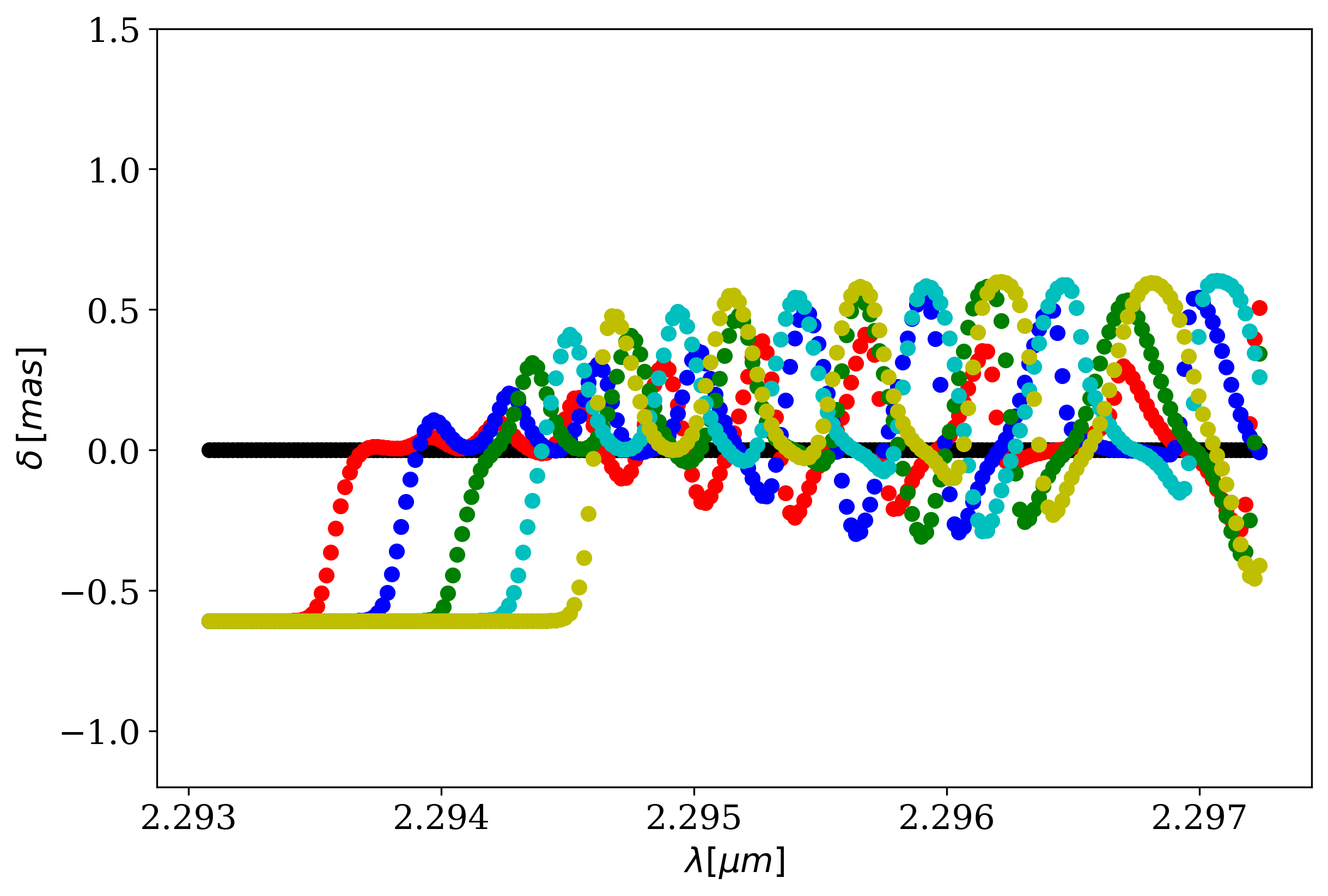}
    \caption{Same as Fig.\,\ref{fig:shifts_h_res_diff_Vrot}, but for displacements along an arbitrary PA of 135\degr.}
    \label{fig:shifts_PA130_h_res_diff_Vrot}
\end{figure*}

\begin{figure*}
    \centering
    \includegraphics[width=1.0\textwidth]{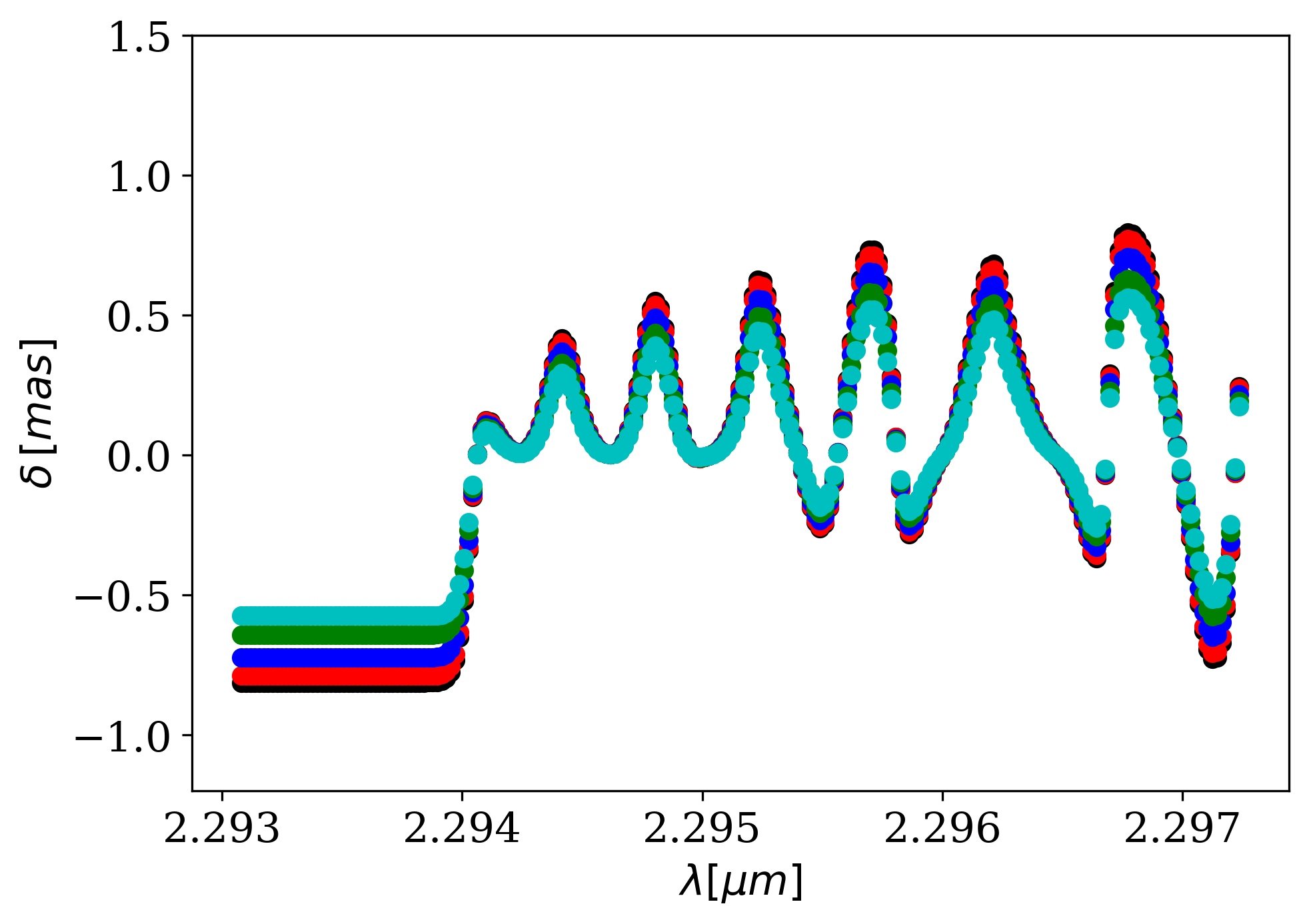}
    \caption{
    Synthetic displacements across the first CO bandhead at R=50\,000 along an aribitrary PA of 135\degr. The CO is emitted in a ring  in Keplerian rotation 0.2\,au across and with \vsini=130\,\kms. Different colours show ring inclinations of 0\degr\ (black), 20\degr\ (red), 40\degr\ (blue), 60\degr\ (green), and 90\degr\ (cyan).
    }
    \label{fig:shifts_h_res_inc}
\end{figure*}

\begin{figure*}
    \centering
    \includegraphics[width=1.0\textwidth]{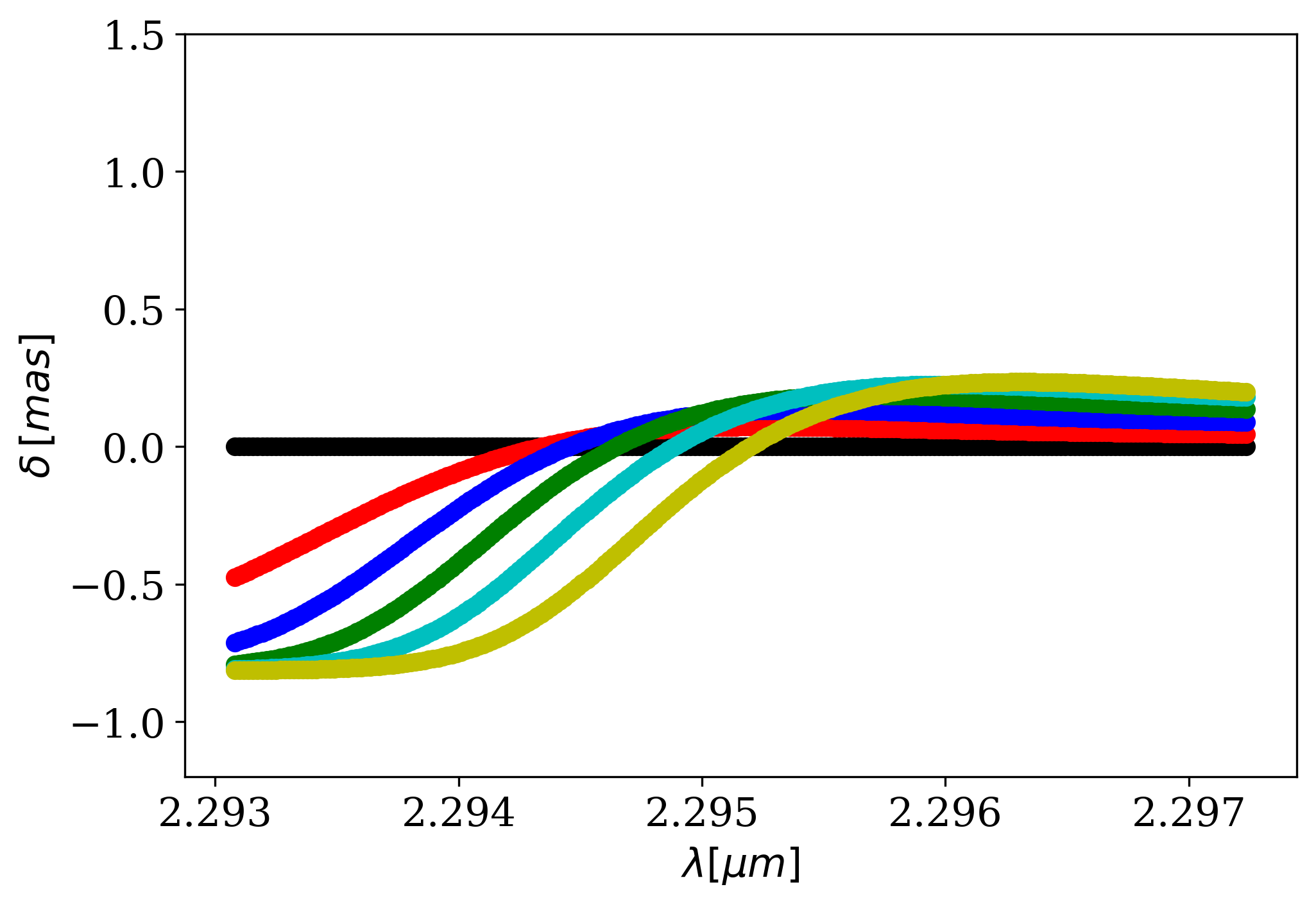}
    \caption{Same as Fig.\,\ref{fig:shifts_h_res_diff_Vrot}, but for R=4000.}
    \label{fig:shifts_P180_R4000_diff_Vrot}
\end{figure*}

\begin{figure*}
    \centering
    \includegraphics[width=1.0\textwidth]{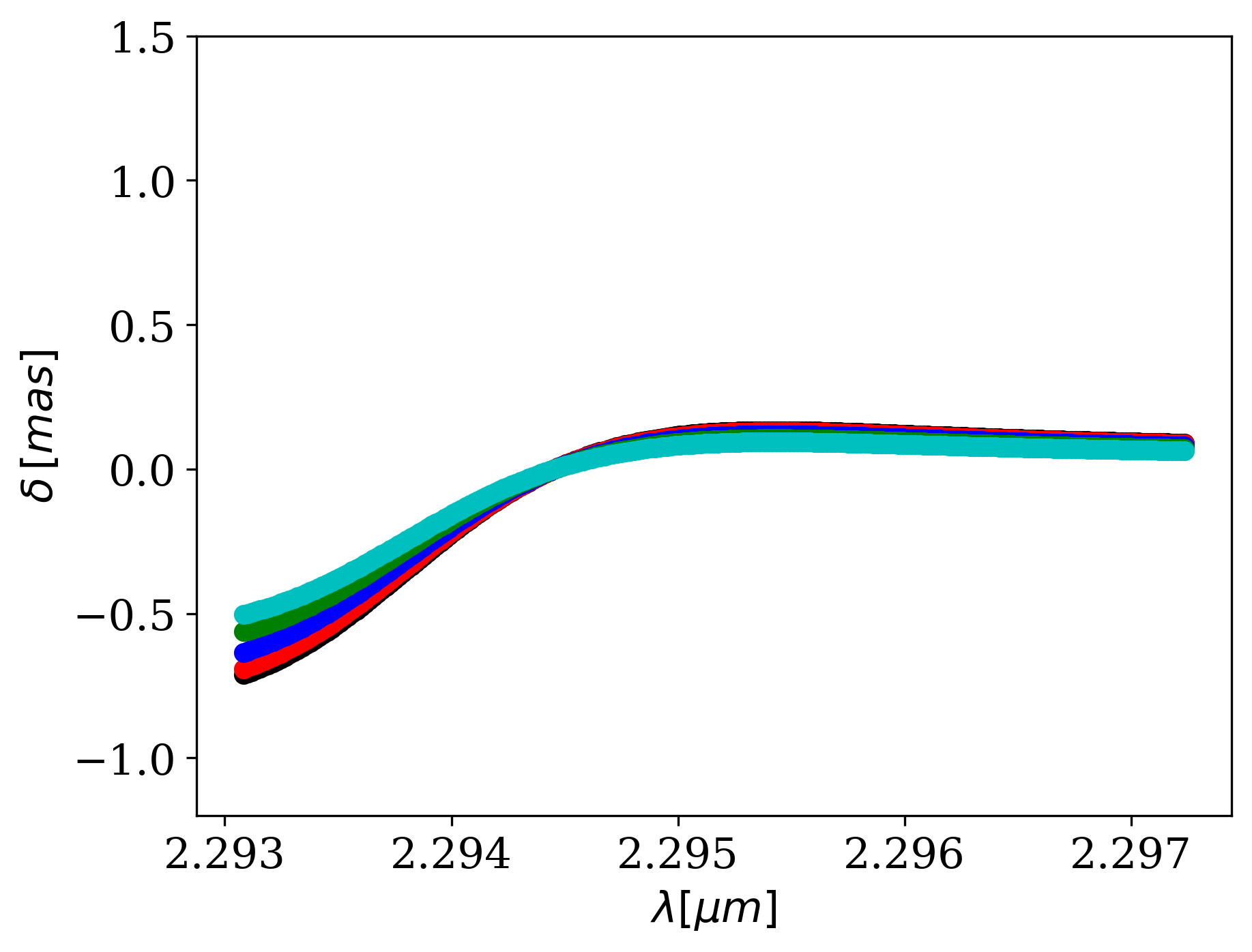}
    \caption{Same as Fig.\,\ref{fig:shifts_h_res_inc}, but for R=4000.}
    \label{fig:shifts_P135_R4000_diff_inc}
\end{figure*}

 The main displacements are always observed at the most strongly blue-shifted wavelength, as the mixture of individual J-components at the peak and red-shifted tails of the bandheads results in small to zero photo-centre shifts. In an inclined ring in Keplerian rotation, the maximum and minimum displacements are achieved along the major and minor axis, respectively, reaching intermediate values.  
Along any fixed PA, the CO line displacement is independent of $v_{rot}$, and it reaches its maximum value along the major axis (Figs.\,\ref{fig:shifts_h_res_diff_Vrot}, \ref{fig:shifts_PA130_h_res_diff_Vrot}, and \ref{fig:shifts_P180_R4000_diff_Vrot}). By probing the line displacements along different PAs, an estimate of $v_{rot}$ can be found.  
On the other hand, for PAs other than that of the ring major axis, the line displacement depends on the ring inclination (see Figs.\,\ref{fig:shifts_h_res_inc} and \ref{fig:shifts_P135_R4000_diff_inc}). This can be used to constrain the ring inclination by comparing the observed and synthetic line displacements along different PAs, assuming different values of $i$.

\begin{figure*}
    \centering
    \includegraphics[width=1.0\textwidth]{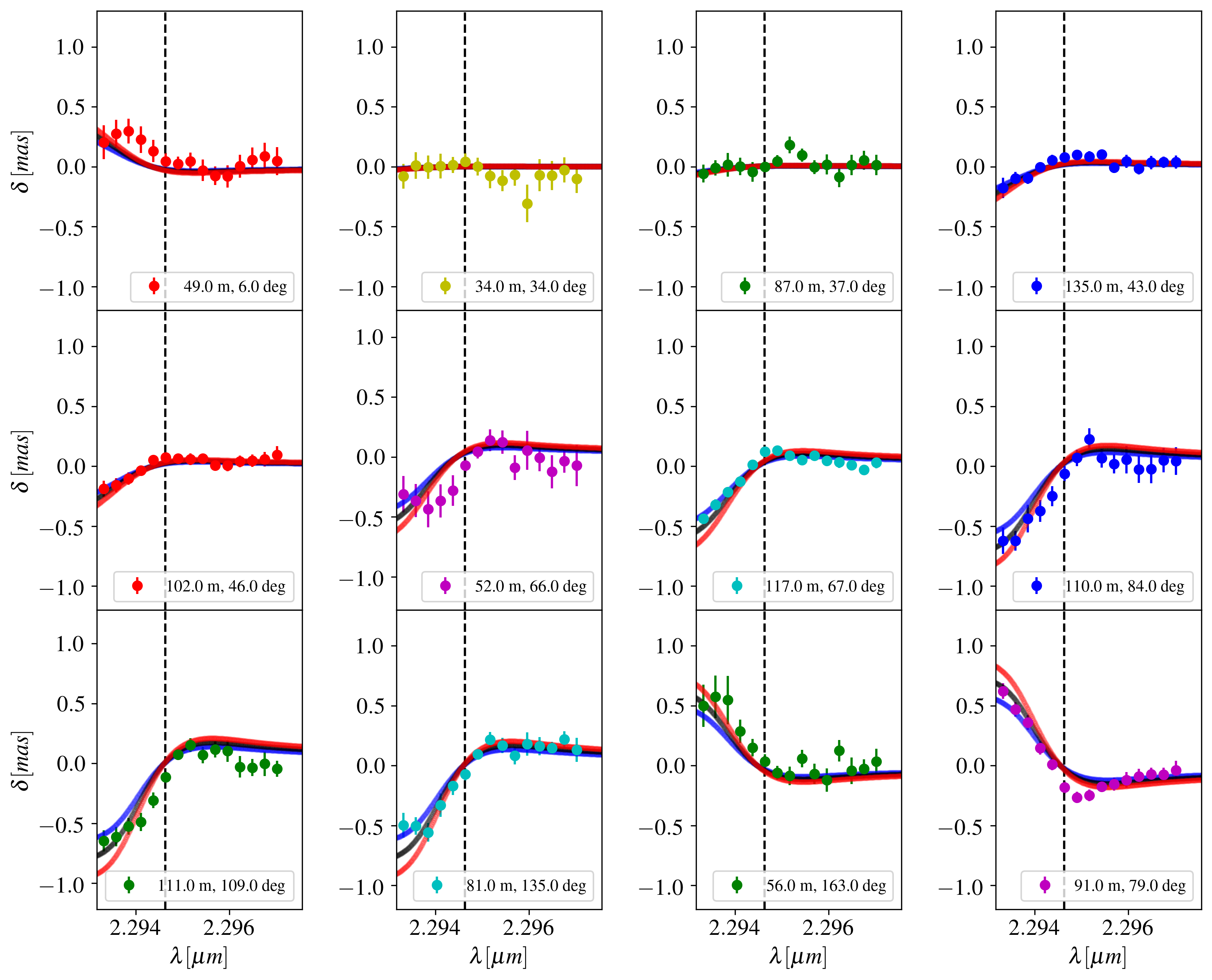}
    \caption{Comparison between observed displacements (coloured dots with error bars) of the first bandhead and the predictions from ring models at R=4000, $i$=70\degr, and radius 0.08\,au (solid blue line), 0.1\,au ( solid black line) and 0.12\,au (solid red line).}
    \label{fig:shifts_diff_Radius}
\end{figure*}

Similarly, the line displacement depends on the size of the emitting region. As before, the main displacements are observed for the most strongly blue-shifted wavelengths, especially along the major axis PA. For fixed values of $i$, and $v_{rot}$, this can be used to derive the size of the CO emitting region (see Fig.\,\ref{fig:shifts_diff_Radius}).

\begin{figure*}
    \centering
    \includegraphics[width=1.0\textwidth]{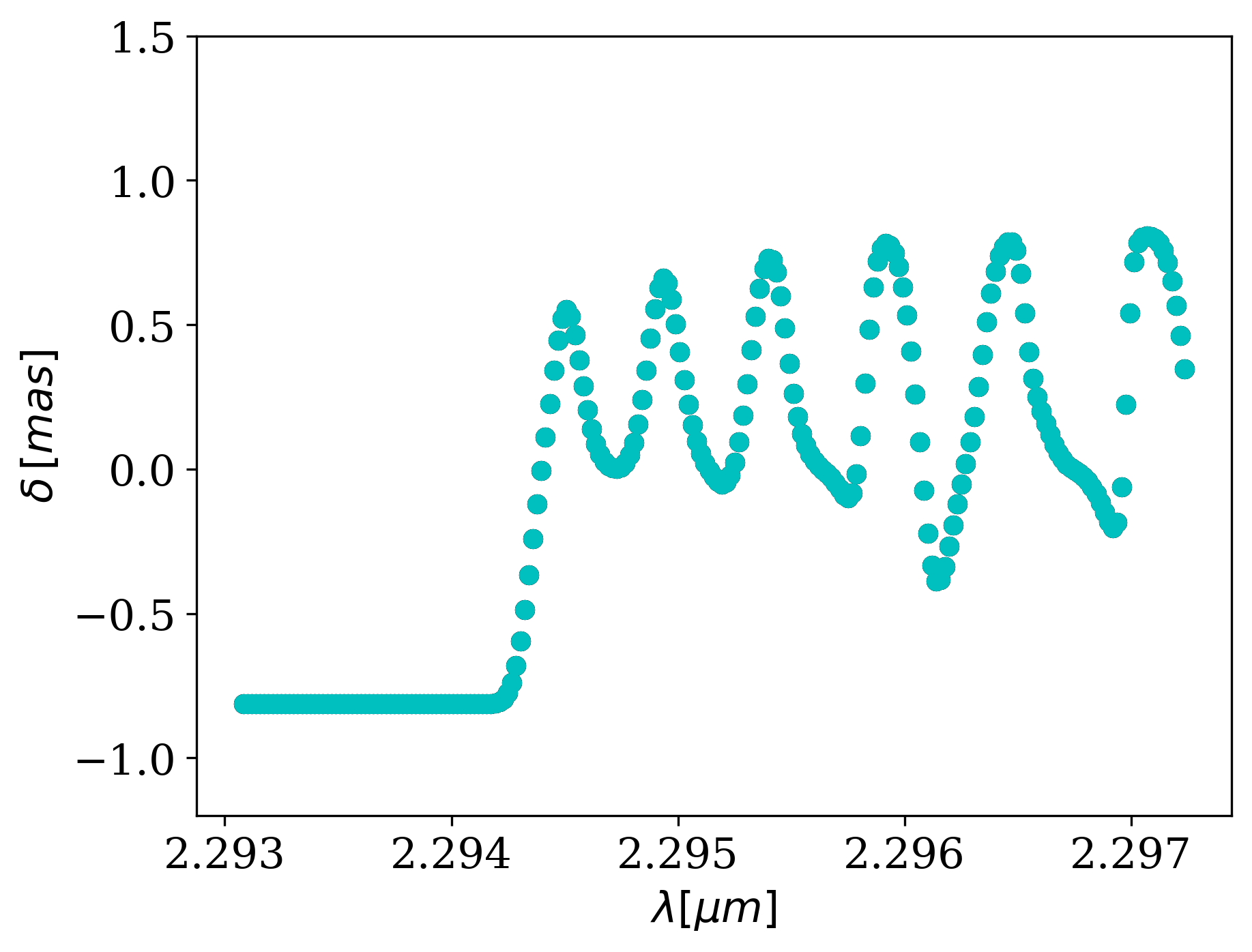}
    \caption{
    Synthetic displacements across the first CO bandhead at R=50\,000 along the major axis (PA=180\degr). The CO is emitted in a ring  in Keplerian rotation 0.2\,au across and with \vsini=130\,\kms. Different colours show ring inclinations ranging from 0\degr\ to 90\degr. The displacement is independent on the inclination value.
    }
    \label{fig:shifts_PA180_h_res_inc}
\end{figure*}

 \end{appendix}

\end{document}